\documentstyle[twoside,fleqn,espcrc2,epsf,rawfonts,oldlfont]{article}

\newcommand{\AmS}{{\protect\the\textfont2
  A\kern-.1667em\lower.5ex\hbox{M}\kern-.125emS}}

\newcommand{\no}{\nonumber \\ }
\newcommand{\se}{self-en\-er\-gy}
\newcommand{\ses}{self-en\-er\-gies}
\newcommand{\fa}{{\em FeynArts\/}}
\newcommand{\fc}{{\em FeynCalc\/}}
\newcommand{\two}{{\em TwoCalc\/}}

\newcommand{\mpar}[1]{{\marginpar{\hbadness10000%
                      \sloppy\hfuzz10pt\boldmath\bf#1}}%
                      \typeout{marginpar: #1}\ignorespaces}
\def\beq{\begin{equation}}
\def\eeq{\end{equation}}
\def\beqar{\begin{eqnarray}}
\def\eeqar{\end{eqnarray}}
\def\barr#1{\begin{array}{#1}}
\def\earr{\end{array}}
\def\bfi{\begin{figure}}
\def\efi{\end{figure}}
\def\btab{\begin{table}}
\def\etab{\end{table}}
\def\bce{\begin{center}}
\def\ece{\end{center}}

\def\text{\textstyle}

\def\al{\alpha}

\def\Ga{\Gamma}

\def\draftdate{\relax}
\def\mda{\relax}
\def\mua{\relax}
\def\mla{\relax}
\def\draft{
\def\thtystars{******************************}
\def\sixtystars{\thtystars\thtystars}
\typeout{}
\typeout{\sixtystars**}
\typeout{* Draft mode!
         For final version remove \protect\draft\space in source
file *}
\typeout{\sixtystars**}
\typeout{}
\def\draftdate{\today}
\def\mda{\mpar{$\downarrow$}}
\def\mua{\mpar{$\uparrow$}}
\def\mla{\marginpar[\boldmath\hfil$\rightarrow$\hfil]%
                   {\boldmath\hfil$\leftarrow $\hfil}%
                    \typeout{marginpar:
$\leftrightarrow$}\ignorespaces}
\overfullrule 5pt
\marginparwidth 19mm 
}

\def\Vhat{\hat V}
\def\xiQ{\xi_Q}

\begin{document}

\thispagestyle{empty}
\null
\hfill INLO-PUB-11/94\\
\mbox{} \hfill hep-ph/9406404
\vskip 1cm
\vfill
\begin{center}
{\Large \bf 
\boldmath{Calculation of Two-Loop Self-Energies in the\\
electroweak Standard Model\footnote{To appear in Nucl.~Phys.~B
(Proceedings Supplements)}
}
\par} \vskip 2.5em
{\large
{\sc S.~Bauberger, M.~B\"ohm, G.~Weiglein%
\footnote{E-mail address: weiglein@vax.rz.uni-wuerzburg.d400.de}
} \\[1ex]
{\normalsize \it Institut f\"ur Theoretische Physik,
Universit\"at W\"urzburg\\
Am Hubland, D-97074 W\"urzburg, Germany}
\\[2ex]
{\sc F.A.~Berends, M.~Buza}\\[1ex]
{\normalsize \it Instituut-Lorentz, University of Leiden\\
P.O.B.~9506, 2300 RA Leiden, The Netherlands}
\par} \vskip 1em
\end{center} \par
\vskip 4cm 
\vfil
{\bf Abstract:} \par
Motivated by the results of the electroweak precision
experiments, studies of two-loop \se\ Feynman diagrams are
performed. An algebraic method for the reduction of
all two-loop \ses\ to a set of standard scalar integrals is
presented. 
The gauge dependence of the \ses\ is discussed and
an  extension of the
pinch technique to the two-loop level is worked out.
It is shown to yield a special case of the background-field
method 
which provides a general framework for deriving
Green functions with desirable theoretical properties. 
The massive scalar integrals of \se\ type are expressed in terms
of generalized multivariable hypergeometric functions. The
imaginary parts of these integrals yield complete elliptic
integrals. Finally, one-dimensional integral representations
with elementary integrands are derived which are well suited for
numerical evaluation.

\par
\vskip 1cm 
\noindent INLO-PUB-11/94 \par
\vskip .15mm
\noindent June 1994 \par
\null
\setcounter{page}{0}
\setcounter{footnote}{0}
\clearpage

\title{Calculation of two-loop self-energies in the electroweak
Standard Model\thanks{Combined talks, presented by F.A.~Berends,
M.~B\"ohm and G.~Weiglein at the Zeuthen Workshop on Elementary Particle
Theory --- Physics at LEP200 and Beyond, held at Teupitz, Germany, April
10--15 1994}
\thanks{Research supported by EU contract
CHRX-CT-92-0004, Stichting FOM and Deutsche
Forschungsgemeinschaft}}

\author{S.~Bauberger,\address{Institut f\"ur Theoretische Physik,
        Universit\"at W\"urzburg, Am Hubland, D-97074
        W\"urzburg, Germany}%
        F.A.~Berends,\address{Instituut-Lorentz, University of
        Leiden, P.O.B.~9506, 2300 RA Leiden, The Netherlands}%
        M.~B\"ohm,$^{\mathrm{a}}$
        M.~Buza$^{\mathrm{b}}$
        and
        G.~Weiglein$^{\mathrm{a}}$}


\begin{abstract}
Motivated by the results of the electroweak precision
experiments, studies of two-loop \se\ Feynman diagrams are
performed. An algebraic method for the reduction of
all two-loop \ses\ to a set of standard scalar integrals is
presented. 
The gauge dependence of the \ses\ is discussed and
an  extension of the
pinch technique to the two-loop level is worked out.
It is shown to yield a special case of the background-field method 
which provides a general framework for deriving
Green functions with desirable theoretical properties. 
The massive scalar integrals of \se\ type are expressed in terms
of generalized multivariable hypergeometric functions. The
imaginary parts of these integrals yield complete elliptic
integrals. Finally, one-dimensional integral representations
with elementary integrands are derived which are well suited for
numerical evaluation.
\end{abstract}
\maketitle

\input prepictex
\input pictex
\input postpictex

\section{INTRODUCTION}

The beautiful results of the LEP experiments~\cite{Blondel} have shown
that the electroweak theory has a predictive power like QED several
decades ago. It is to be expected that eventually the electroweak
theory will provide high precision predictions for many experiments
in the present and near future. One may in particular think in this
respect of the measurement of the $Z$ mass and width, a better
determination of the $W$ mass and a 
direct indication of the top mass.

These developments will require in the future two-loop calculations
in the electroweak theory. 
In the study of the two-loop contributions the self-energy
diagrams play a central role.
The present paper reviews the problems
one encounters when considering self-energies in the Standard Model
(SM) and discusses recently developed methods to tackle them.

The choice to treat first the self-energies is a logical one.
They are the simplest two-loop diagrams which yield a universal
contribution to all two-loop processes. Moreover, from one-loop
calculations we know their importance. 
Several results for two-loop \ses\ are known, involving 
however approximations. They concern the
limiting cases of a heavy fermion doublet~\cite{vdBH}, a heavy top
quark~\cite{barb2} and a large Higgs mass~\cite{vdBV}. 
However, in the electroweak theory also
the $W$ and $Z$ bosons have a non-negligible mass, and one is 
in general faced
with two-loop self-energies containing massive propagators.

Let us summarize the problems which arise when evaluating 
two-loop self-energies.
The first problem is the plethora of Feynman diagrams
contributing, the second one the evaluation of tensor integrals in
terms of scalar integrals and the third one the calculation of the
scalar integrals themselves.
When discussing the corrections due to two-loop \ses, one 
furthermore has to deal with the problem that these
contributions are in general gauge dependent.

In order to handle the large number of Feynman diagrams
needed for the evaluation of two-loop \ses, an
algebraic approach is chosen allowing for a high degree of
automatization.
It involves a method for the decomposition of tensor integrals.
Whereas for one-loop diagrams this is a well-known 
procedure~\cite{pass}, it
was only recently developed for two-loop \ses~\cite{Red}.
In this way all two-loop \ses\ can be reduced to a set of
standard scalar integrals. The task of evaluating any two-loop
self-energy involving in general several thousands of Feynman
amplitudes is therefore reduced to the problem of calculating
four different types of two-loop scalar integrals.
As an example, we treat light fermion contributions to the 
\se\ of the $Z$-boson.
The results given in a minimal basis of standard integrals allow
to study the gauge dependence of the considered quantities
directly at the algebraic level. 

At one-loop order the \ses\ are frequently used as building
blocks to define running couplings or to
parametrize electroweak radiative corrections. The need for
\ses\ as building blocks with suitable theoretical properties
seems to be even more important at the two-loop level, where so
far no calculation of a complete process exists.
In the one-loop applications it was found that the \ses\
evaluated in the class of $R_\xi$-gauges are not adequate 
as building blocks due to
their gauge dependence and unsatisfactory high energy and UV
behavior. Many proposals to modify these self-energies aimed
on eliminating their gauge parameter dependence. In particular,
the pinch technique (PT)~\cite{pinch} was found to yield
results with decent properties. Recently it was shown~\cite{bgf}
that the background-field method (BFM)~\cite{Ab81}
offers a wider framework in
which Green functions possessing desirable theoretical properties 
are directly derived from a gauge invariant effective
action. In this paper we work out an extension of the
PT to the two-loop level and show that it corresponds to a
special case of the BFM results.

The evaluation of the two-loop scalar integrals is in general
needed for non-negligible masses and an arbitrary dimension $D$.
When expanding around
$\delta=(4-D)/2 \sim 0$, one gets in 
these cases not anymore
results in terms of (poly) logarithms. To perform nevertheless an
evaluation of the massive two-loop diagrams, in essence two approaches
are followed: expansions for small and large external momenta and
numerical integration. In both strategies this
paper presents
recent new results. On one hand, the use of multi-variable generalized
hypergeometric functions, i.e. multiple series, is a new development.
On the other hand, one-dimensional integral representations are derived,
which are a good alternative to the existing two-dimensional integrals.
It turns out that the derivation of one-dimensional integrals
is possible for all two-loop diagrams which contain a one-loop
self-energy insertion.

For the imaginary parts of the scalar integrals simple 
analytic results in terms of complete elliptic
integrals are derived.

The outline of the paper is as follows.

The algebraic method for reducing two-loop \ses\ to standard
scalar integrals is described in section 2,
whereas in section 3 the question of gauge invariance in the
framework of PT and BFM is discussed. 
The next section deals with an analytic approach
to scalar self-energy integrals leading to generalized hypergeometric
functions. Section 5 focuses on the imaginary part of the scalar
integrals, which are then expressed in terms of elliptic integrals. 
The one-dimensional integral representations are derived in section 6.
In section 7 we give our conclusions and an outlook.

\section{ALGEBRAIC CALCULATION OF TWO-LOOP SELF-ENERGIES}

\subsection{Classification}

\begin{figure*}
\begin{picture}(450,20)
\put(-44,-520){\includegraphics{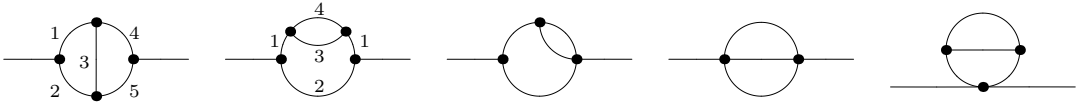}}
\put(42, 8){${\scriptstyle 1}$}
\put(72, 8){${\scriptstyle 4}$}
\put(53,-3){${\scriptstyle 3}$}
\put(42,-14){${\scriptstyle 2}$}
\put(72,-14){${\scriptstyle 5}$}
\put(125, 5){${\scriptstyle 1}$}
\put(159, 5){${\scriptstyle 1}$}
\put(142, -12){${\scriptstyle 2}$}
\put(142, -1){${\scriptstyle 3}$}
\put(142, 17){${\scriptstyle 4}$}
\end{picture}
\caption{One-particle irreducible topologies of two-loop \ses.}
\label{fig:top}
\end{figure*}

As the first step in the evaluation of two-loop \ses, the
relevant Feynman diagrams have to be generated. These can be
classified according to their topologies.
In fig.~\ref{fig:top} some examples of irreducible two-loop 
self-energy topologies are shown.
A complete set including also tadpoles and topologies which 
factorize into one-loop contributions is given in~\cite{Red}.
Inserting fields into the topologies in all possible ways and
applying the Feynman rules yields all relevant Feynman
amplitudes. The generation of the Feynman diagrams and amplitudes is
carried out automatically with the program \fa\ \cite{fea}.

In the following we will use the shorthand notations 
\begin{displaymath}
\langle \langle \ldots \rangle \rangle = 
 \int \frac{d^D q_1}{i \pi ^2 (2 \pi \mu)^{D - 4}} 
 \int \frac{d^D q_2}{i \pi ^2 (2 \pi \mu)^{D - 4}} ( \ldots ) ,
\end{displaymath}
where $q_1$ and $q_2$ are the integration momenta of the loop
integrals and $\mu $ is an arbitrary reference mass,
and
\begin{displaymath}
d_{i j \ldots l} =
\frac{1}{\Bigl[k_i^2 - m_i^2 \Bigr]
\Bigl[k_j^2 - m_j^2 \Bigr] \cdots \Bigl[k_{l}^2 - m_l^2
\Bigr]}.
\end{displaymath}
Here $k_{i}$ is the momentum of the $i$-th propagator and $m_i$ its
mass. In the most general case corresponding to the first topology in
fig.~\ref{fig:top} five different propagators appear. 
Their momenta are related to the integration momenta $q_1$ and
$q_2$ and the external momentum $p$ via
\begin{eqnarray}
k_1 &=& q_1, \; k_2 = q_1 + p, \; k_3 = q_2 - q_1, \; k_4 = q_2,
\; \no
k_5 &=& q_2 + p. \label{eq:momenta}
\end{eqnarray}

A general Feynman amplitude takes the form 
\begin{equation}
\langle \langle f(k_i, m_i, D) \, d_{1 2 \ldots  l} 
\rangle \rangle , 
\label{eq:den}
\end{equation}
where $f(k_i, m_i, D)$ is an
expression depending on the momenta, the particle
masses and the space-time dimension $D$ 
and containing the whole Lorentz and Dirac structure.

As will be discussed in the next section, all two-loop \ses\ can 
be reduced to a class of standard scalar integrals which we call
$T$-integrals
\begin{eqnarray}
\lefteqn{T_{i_1 i_2 \ldots i_l}
(p^2;m_1^2,m_2^2,\ldots,m_{l}^2 ) = } \no
\lefteqn{
\; \; = \langle \langle
\frac{1}{ \Bigl[k_{i_1}^2 - m_1^2 \Bigr] \Bigl[k_{i_2}^2 -
m_2^2 \Bigr] \cdots \Bigl[k_{i_l}^2 - m_l^2 \Bigr] }
\rangle \rangle , }  \label{eq:Tint}
\end{eqnarray}
where the denominator is of the same form as in the Feynman amplitude.
The double index notation $T_{i_1 i_2 \ldots i_l}$ is used here
to indicate that the
indices of the $T$-integrals refer to the corresponding
momenta $k_{i_1}, k_{i_2}, \ldots k_{i_l}$.
In the following the momentum $p^2$ and the masses will only
be explicitly written as arguments if confusion is possible.

The topologies depicted in fig.~\ref{fig:top} correspond to the
scalar integrals $T_{12345}, T_{11234}, T_{1234}, T_{234}$ and
$T_{1134}$, respectively. The analytical expression for
$T_{11234}$ can be obtained from $T_{1234}$ by 
partial fractioning or taking the
derivative with respect to $m_1^2$. 
Other integrals with higher powers of propagators are treated
in the same way.
For the general case one therefore needs to evaluate only
four different types of two-loop scalar integrals.

\subsection{Reduction to standard scalar integrals}

It is convenient to begin with a decomposition into Lorentz
scalars. For the gauge boson self-energies it reads 
\begin{displaymath}
\Sigma _{\mu \nu}(p) = \Bigl(-g_{\mu \nu }+ \frac {p_{\mu}
p_{\nu}}{p^2} \Bigr) \Sigma _{T}(p^2) - \frac {p_{\mu}
p_{\nu}}{p^2} \Sigma _{L}(p^2) ,
\end{displaymath}
from which the transverse part $\Sigma _{T}(p^2)$ 
and the
longitudinal part $\Sigma _{L}(p^2)$ can
be obtained
\begin{eqnarray*}
\Sigma _{T}(p^2) &=& \frac{1}{D-1}
\Bigl(-g^{\mu \nu } + \frac {p^{\mu} p^{\nu}}{p^2} \Bigr)
\Sigma _{\mu \nu}(p)  ; \no
\Sigma _{L}(p^2) &=& - \frac {p^{\mu} p^{\nu}}{p^2}
\Sigma _{\mu \nu}(p) .
\end{eqnarray*}
For all other types of \ses\ scalar quantities can be extracted
in a similar way~\cite{Red}.

The contraction of Lorentz indices, reduction of the Dirac
algebra and evaluation of Dirac traces can be worked out like in
the one-loop case, e.g.~with the program \fc\ \cite{fec}.
Since we deal with scalar quantities, this results in
scalar products of momenta $(k_i \cdot k_j),$ $\, (k_i \cdot
p),$ $\, p^2$ in the
numerator of the Feynman amplitude. The
denominator is unchanged. We now implicitly use momentum
conservation and express all scalar products as sums of momentum
squares, e.g.
\begin{equation}
(k_1 \cdot p) = \frac {1}{2} (k_2^2 - k_1^2 - p^2)  .
\end{equation}
Subsequently all $k_i^2$ appearing both in the numerator and
the denominator are canceled via
\begin{equation}
k_i^2 = (k_i^2 - m_i^2) + m_i^2  .
\end{equation}

In general there remain $k_i^2$ in the numerator which cannot be
canceled, e.g.~for 
\begin{equation}
I = \langle \langle k_5^2 d_{1 2 3 4} \rangle \rangle .
\end{equation}
We proceed in this case by writing
\begin{equation}
k_5^2 = (k_4^2 - m_4^2) + (m_4^2 + p^2) + 2 (p \cdot k_4) .
\end{equation}
The first term can be canceled with the appropriate propagator 
factor in the denominator. 
The second term contains no integration momenta.
We therefore focus on the integral
\begin{equation}
p_{\mu} S_{1 2 3 4}^{4, \; \mu} = p_{\mu} \langle \langle k_{4}
^{\mu}  d_{1 2 3 4} \rangle \rangle 
\label{eq:newtens}
\end{equation}
and perform a tensor decomposition for $S_{1 2 3 4}^{4, \; \mu}$.

In contrast to the one-loop case the
decomposition with respect to the external momentum, i.e.~the
ansatz~\cite{pass} 
\begin{equation}
S_{1 2 3 4} ^{4 , \; \mu} = p^{\mu} S(p^2) ,
\label{eq:pas}
\end{equation}
does not lead to simpler integrals here. In order to determine
the scalar quantity $S(p^2)$, one has to contract with $p_{\mu}$
yielding terms in the numerator which cannot be canceled.
This fact is due to the structure of the 
topology associated with $S_{1 2 3 4} ^{4 , \; \mu}$  (the third
topology in fig.~\ref{fig:top}) which contains a four-vertex with 
three inner lines. This is a typical feature of two-loop
topologies not present at the one-loop level.

Instead, we perform a decomposition with respect to a
subloop. We write
\begin{eqnarray}
\lefteqn{\langle k_{4}^{\mu} d_{3 4} \rangle \equiv } \no
\lefteqn{ \equiv 
\int \frac{d^D q_2}{i
\pi ^2 (2 \pi \mu)^{D - 4}} \frac{q_2^{\mu}}{\Bigl[(q_2 - k_1)^2
- m_3^2 \Bigr] \Bigl[q_2^2 - m_4^2\Bigr]}}  \no
\lefteqn{= k_1^{\mu} s(k_1^2) ,}
\end{eqnarray}
where the last equality follows from the fact that the tensor integral
depends only on $k_1^{\mu}$. The quantity $s(k_1^2)$ is given by
\begin{equation}
s(k_1^2) = \frac{1}{k_1^2} \langle (k_1 \cdot k_4) d_{3 4}
\rangle = \langle (k_1 \cdot k_4) d_{1'34} \rangle .
\end{equation}
The factor $1/k_1^2$ can be written as a massless 
propagator, which we have
indicated with a prime at the corresponding index in
$d_{1'34}$. Inserting this into~(\ref{eq:newtens}) yields
\begin{equation}
p_{\mu} S_{1 2 3 4}^{4, \; \mu} = \langle \langle (p \cdot k_1) (k_1
\cdot k_4) d_{1'1234} \rangle \rangle .
\end{equation}
The scalar products in the numerator of this integral can now be
expressed in terms of squared momenta which can be canceled.
Therefore all momentum dependent terms have been removed from
the numerator of the Feynman integral while its denominator has
retained its original structure. It can further be simplified by
performing a partial fractioning for the propagator factors
$d_{1'1}$ carrying the same momentum.

It can be shown~\cite{Red} that in a similar way all possible integrals 
appearing in calculations of general two-loop
\ses\ can be reduced to $T$-integrals.

The result of the algebraic calculation in general contains several
scalar integrals depending on different arguments. 
These are not necessarily
independent of each other, i.e.~there exist relations
\begin{equation}
c^1 T^1 + c^2 T^2 + \ldots + c^n T^n  = 0 
\label{eq:relTn}
\end{equation}
where $c^1, \ldots , c^n$ are polynomials
in $p^2$. We use these relations together with the symmetry
properties of the $T$-integrals to eliminate as many integrals as
possible. This leads to a 
minimal set of scalar integrals which are algebraically
independent of
each other. The result expressed in this minimal basis
is very transparent and 
directly displays certain properties of the considered
quantity. 

For example, Slavnov-Taylor identities can be checked
at a purely algebraic level without having to use any
explicit expression for the $T$-integrals. When adding up the
results for the relevant amplitudes, the coefficient of every standard
integral exactly adds up to zero.
This has been demonstrated in~\cite{Red} for Slavnov-Taylor
identities involving several thousands of Feynman amplitudes.

The algorithms outlined here have been implemented into the
program \two~\cite{two} which carries out the algebraic 
calculation fully automatically.

\subsection{Results for the Standard Model}

\begin{figure*}[t]
\begin{picture}(350,80)
\put(-45,-470){\includegraphics{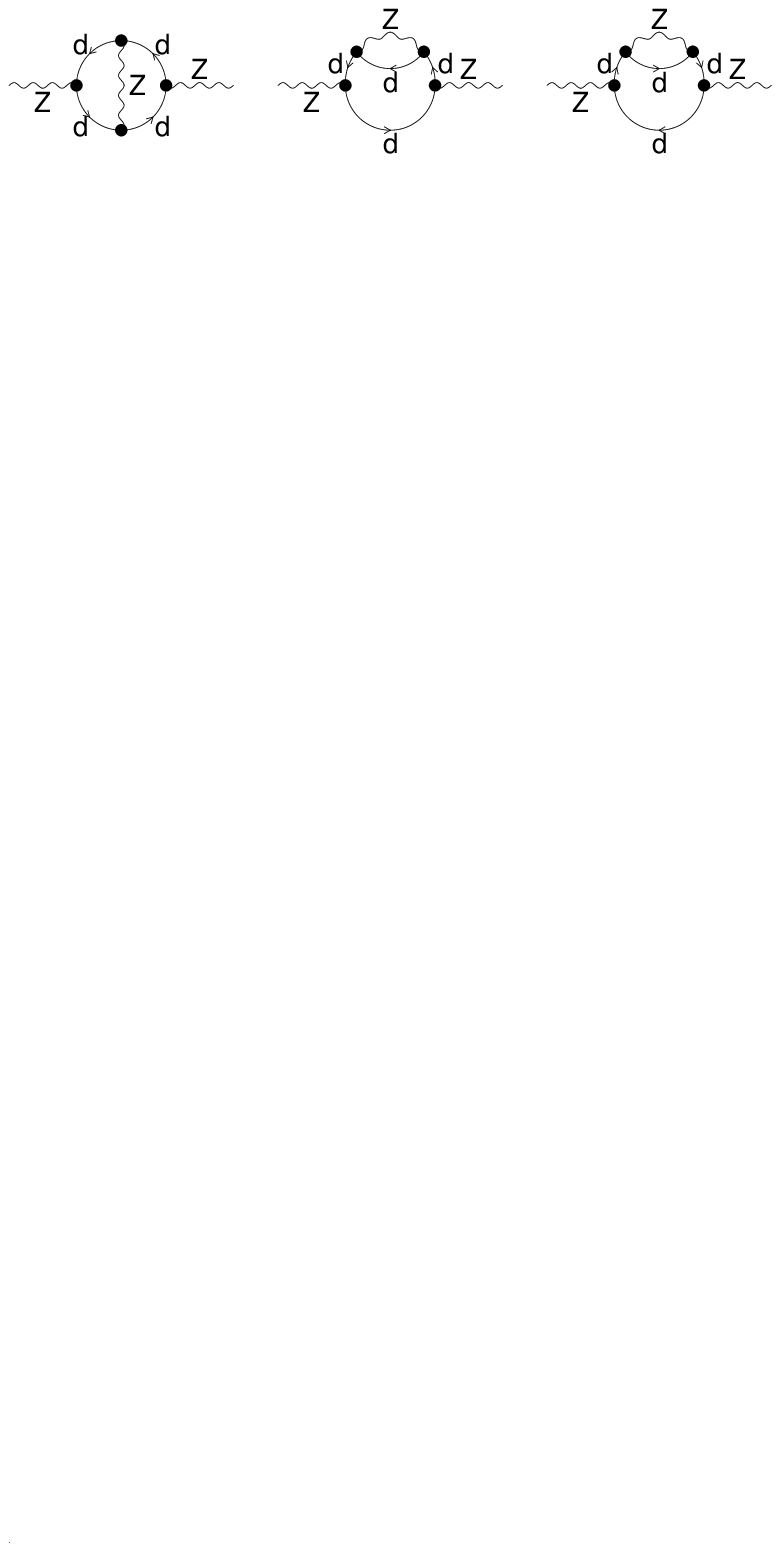}}
\put(-45,-530){\includegraphics{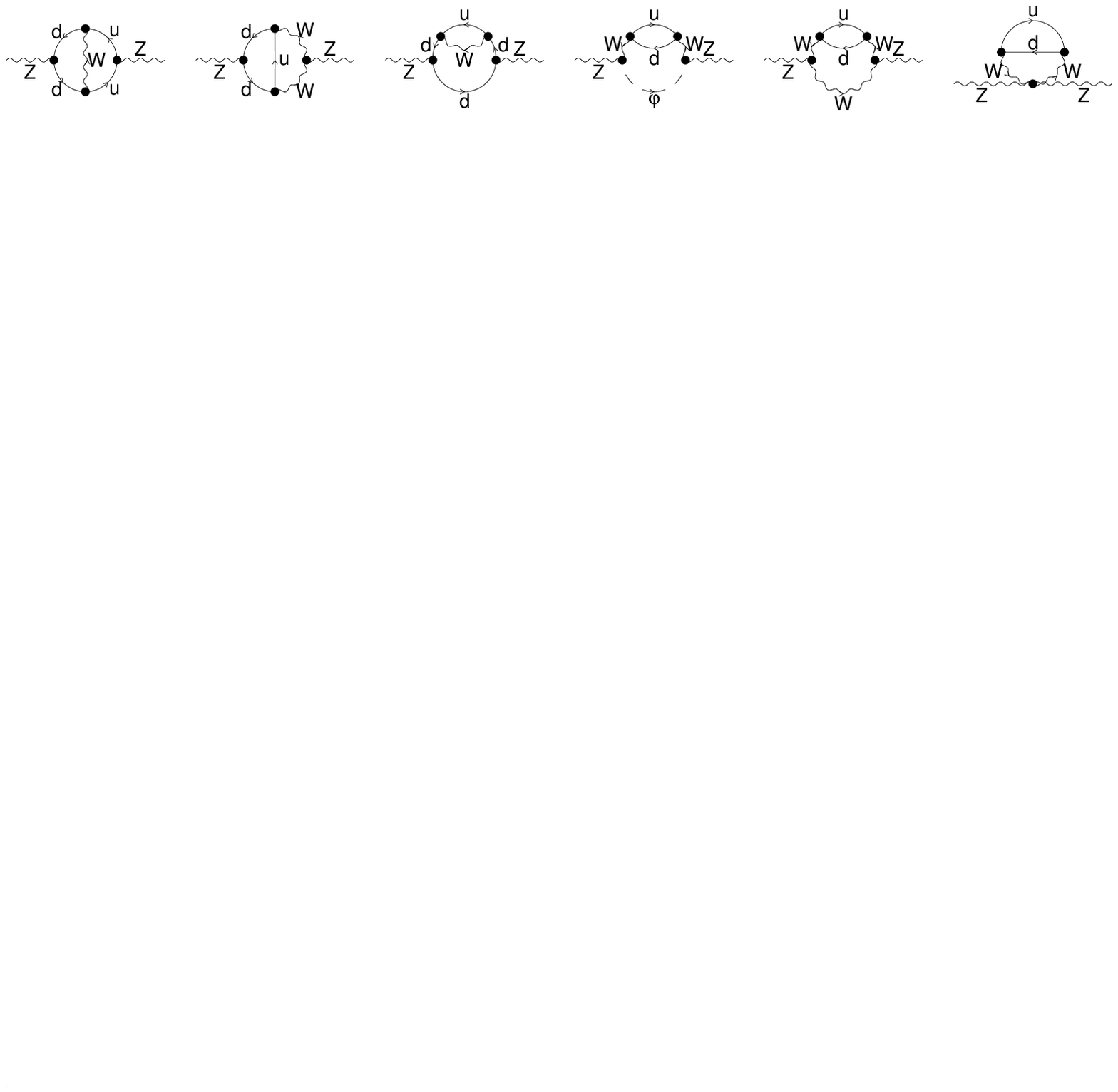}}
\put(221,39){$(a)$}
\put(221,-21){$(b)$}
\end{picture}
\caption{Light fermion contributions to the two-loop self-energy
of the $Z$-boson.}
\label{fig:diagSM}
\vspace{-0.2 cm}
\end{figure*}

Another feature which can directly be read off from the
algebraic result is its gauge dependence. We work in a general
$R_{\xi}$-gauge specified by one gauge parameter $\xi_i$ ($i =
\gamma , Z, W$) for each vector boson 
and write the lowest-order gauge boson propagators as
\begin{equation}
 \Delta^{i}_{\mu \nu} = 
\frac{-i g_{\mu \nu }}{ \Bigl[ k^2 - m_i^2
\Bigr]}
 + \frac{i (1 - 1/ \xi_i )
 k_{\mu} k_{\nu}}{ \Bigl[ k^2 - m_i^2 / \xi_i \Bigr]  \Bigl[ k^2
-
 m_i^2 \Bigr] } . \label{eq:gabopr}
\end{equation}
As a simple example, we consider two sets of light fermion graphs
contributing to the $Z$ \se. They are shown in 
fig.~\ref{fig:diagSM}. A complete listing of 
the light fermion contributions to the gauge
boson \ses\ was given in~\cite{Red}.
All fermions except the top-quark can be treated as light
fermions, i.e.~their masses are small compared to those
of the $Z$- and $W$-bosons and can therefore be neglected.
For definiteness, we choose $d$- and $u$-quark.

We find for the transverse part of the three ``neutral current''
diagrams depicted in fig.~\ref{fig:diagSM}a
\begin{eqnarray}
\lefteqn{\Sigma ^{ Z Z,(2)}_{T, Z}(p^2) = e^4 C_Z \Bigl[
         f_Z(A_0, B_0)} \no
&& + (4 - 6 D + D^2) T_{13'4'}(m_Z^2) \no
&& + (8 - 4 D + D^2) T_{23'4'}(p^2;m_Z^2) \no
&& - (4 - 8 D + D^2) (m_Z^2 + p^2) T_{1'2'34'}(p^2;m_Z^2) \no
&& - D (m_Z^4 + 4 m_Z^2 p^2 - m_Z^2 p^2 D/2 + p^4) \no
&& \times T_{1'2'34'5'}(p^2;m_Z^2) \Bigr]  , \label{eq:ZZ}
\end{eqnarray}
where $C_Z$ is a dimensionless constant 
and $f_Z(A_0, B_0)$ represents a function containing
only scalar one-loop integrals. The result for each single
diagram of fig.~\ref{fig:diagSM}a depends on the gauge parameter
$\xi_Z$. However, as is seen in~(\ref{eq:ZZ}), in the sum of these 
diagrams the gauge parameter has canceled,
i.e.~this contribution is gauge independent within the class of
$R_{\xi}$-gauges.

\begin{figure*}[t]
\begin{picture}(450,40)
\put(-44,-515){\includegraphics{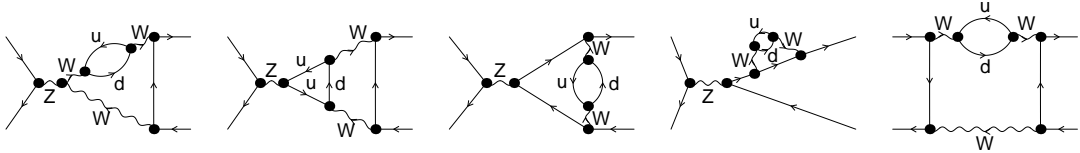}}
\end{picture}
\caption{Two-loop vertex and box graphs containing
propagator-like pinch parts.}
\label{fig:pin}
\vspace{-0.2 cm}
\end{figure*}

Fig.~\ref{fig:diagSM}b represents a set of ``charged current''
diagrams where a $W^{\pm}$ 
is exchanged. We have only drawn 
one diagram of each type. The result for the altogether 15
amplitudes is given by
\begin{eqnarray}
\lefteqn{\Sigma ^{Z Z,(2)}_{T, W^{\pm}}(p^2) = 
         \frac{e^4 C}{m_W^2 m_Z^2} \Bigl[ 
         f(A_0, B_0) } \no
\lefteqn{\quad + g(A_0, B_0; \xi_W) + F(T)} \no
\lefteqn{\quad + 9 (p^2 - m_Z^2) (2 m_W^2 - m_Z^2 - p^2) 
	 \frac{m_W^2}{p^2} G(T; \xi_W) \Bigr] .} \no
&& \label{eq:zW}
\end{eqnarray}
Here $C$ is again a dimensionless constant, 
$f(A_0, B_0)$ and $g(A_0, B_0; \xi_W)$ represent terms which 
only involve scalar one-loop
integrals, and $F(T)$ and $G(T; \xi_W)$ contain the scalar 
two-loop integrals. In contrast to~(\ref{eq:ZZ}), this result is
gauge dependent. For illustration, we only give here the
explicit form of the gauge parameter dependent function $G(T; \xi_W)$:
\begin{eqnarray}
\lefteqn{G(T; \xi_W) =  \frac{1}{\xi_W} (D - 3) 
	 T_{13'4'}(m_W^2)} \no
\lefteqn{\quad + T_{23'4'}(p^2;m_W^2/ \xi_W)} \no
\lefteqn{\quad - 2 \Big[ \Big(\frac{1}{\xi_W} - 1\Big) m_W^2 
         + (3 - 2 D) p^2 \Big]} \no
\lefteqn{\quad \times T_{123'4'}(p^2;m_W^2, m_W^2/ \xi_W)} \no
\lefteqn{\quad + \Big[m_W^4 \Big( \frac{1}{\xi_{W}} - 1 \Big)^2
         - 2 \Big( \frac{1}{\xi_{W}} + 3 - 2 D \Big) m_W^2 p^2
	 } \no
\lefteqn{\quad \; \; \; \; + p^4 \Big]
	 T_{1123'4'}(p^2;m_W^2, m_W^2, m_W^2/ \xi_W) .} \label{eq:GG}
\end{eqnarray}
The gauge parameter appears both in the arguments of the scalar
integrals and in their coefficients. The other terms contributing 
to~(\ref{eq:zW}) were explicitly given in Ref.~\cite{Red}.

The characteristic feature of the result~(\ref{eq:zW}) is the
factor $(p^2 - m_Z^2)$ multiplying the gauge dependent part
$G(T; \xi_W)$ which contains the generic two-loop integrals. 
This reflects the fact
that the pole position of the two-loop propagator is gauge 
independent.
Note that for $p^2 = m_Z^2$ there is still a remaining gauge
dependence in the function $g(A_0, B_0; \xi_W)$. 
It cancels with the gauge dependence of the terms generated by 
inserting one-loop counterterms into the one-loop diagrams. 

\section{GAUGE INVARIANCE OF TWO-LOOP SELF-ENERGIES}


In order to investigate physical effects due to two-loop contributions 
of propagator type, it would be desirable to arrange these
contributions in such a way that they form building blocks 
with suitable theoretical properties. 
At one-loop order, the pinch technique (PT)~\cite{pinch} was
developed to eliminate the gauge parameter dependence by
shifting contributions between different Green functions. 
It was found that the new ``Green functions'' obtained in this 
way fulfill simple Ward identities and in comparison to their
$R_{\xi}$-gauge counterparts possess desirable properties such
as improved IR and UV properties and a decent high-energy
behavior.

However, it was recently shown~\cite{bgf} that the requirement
of gauge parameter independence is not the crucial 
criterion for obtaining
well-behaved Green functions. The background-field method (BFM)
provides a more general framework in which the Green functions
are directly derived from a gauge invariant effective action.
Their desirable theoretical properties are a consequence of
gauge invariance and hold for all values of the quantum gauge
parameter $\xi_Q$. 
In~\cite{bgf} the BFM was applied to QCD and the SM. It was
shown that the BFM includes the PT results as 
the special case $\xiQ = 1$.
Viewed from the framework of the BFM, the PT results therefore are
not gauge independent but correspond to a certain choice of the
quantum gauge parameter $\xi_Q$.

In order to discuss this issue at the two-loop level, we present
here for an example the application of both approaches.
In contrast to the BFM, the PT has so far been restricted to the
one-loop level only. We therefore work out an extension to the
two-loop case.
We treat the ``charged current'' light fermion contributions to
the two-loop $Z$ self-energy. The result in the $R_{\xi}$ gauge
was given in~(\ref{eq:zW}).

\subsection{Pinch technique at the two-loop level}


In the PT, the gauge parameter dependence of \ses\ is canceled by
combining them with propagator-like pieces
extracted from vertex and box diagrams which contribute to a
gauge independent S-matrix element. The relevant types of 
two-loop vertex (containing also the wave function corrections) and
box diagrams for a four-fermion process are drawn in 
fig.~\ref{fig:pin}. 


%
\begin{figure}
\begin{picture}(200, 40)
\put(-8,-425){\includegraphics{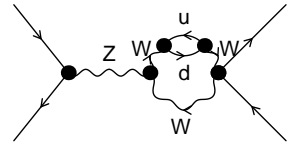}}
\end{picture}
\caption{The pinch part of the first diagram in
fig.~3.}
\label{fig:pinpart}
\vspace{-0.5 cm}
\end{figure}

Parts of these vertex and box graphs 
in which the propagators of the external fermions have been canceled
become pro\-pa\-ga\-tor-like.
Such a ``pinch part'' is depicted in
fig.~\ref{fig:pinpart}. It corresponds to the first diagram in
fig.~\ref{fig:pin}, in which the propagator associated with the
final state fermions has been removed.
The precise definition of the pinch parts a priori is not unique.
In analogy to the one-loop case, we identify the pinch parts as
terms emerging when a longitudinal momentum $k_{\mu}$ of an
elementary vertex or propagator
is contracted with a coupling matrix $\gamma^{\mu}$ of the
external fermion current. This gives rise to an elementary
Ward identity, i.e.~it can be written in terms of
inverse fermion propagators and mass terms
\beq
k_{\mu} \gamma^{\mu} 
= (\not \! p_f + \not \! k
- m_1) - (\not \! p_f - m_2) + m_1 - m_2 ,
\eeq
where $p_f$ is the momentum of the external fermion.
The first term on the right-hand side cancels the fermion propagator 
and therefore gives rise to the pinch part, whereas on-shell
the other terms yield contributions proportional to the
fermion masses.


The $k_{\mu}$-terms originate from the 
longitudinal part of the gauge boson propagators in the $R_{\xi}$-gauge
(see~(\ref{eq:gabopr})) and from the three-gauge-boson vertices. 
Up to group theoretical factors these vertices can be 
decomposed into 
\beqar
\Ga_{\mu \nu \al} &=& \Ga_{\mu \nu \al}^P + \Ga_{\mu \nu \al}^F,
\; \mbox{ where} \no
\Ga_{\mu \nu \al}^P &=& (q + k)_{\nu} g_{\mu \al} 
     + k_{\mu} g_{\nu \al} , \no
\Ga_{\mu \nu \al}^F &=& 2 q_{\mu} g_{\nu \al} - 2 q_{\nu} g_{\mu \al} 
                        - (2 k + q)_{\al} g_{\mu \nu} .
\eeqar
The momenta $(q + k)_{\nu}$ and $k_{\mu}$ in $\Ga_{\mu \nu
\al}^P$ have the same form as those arising from the
longitudinal part of the relevant gauge boson propagators.
Therefore $\Ga_{\mu \nu \al}^P$ also gives rise to pinch parts. 
Note that 
in the first diagram of
fig.~\ref{fig:pin} only the part of $\Ga_{\mu \nu \al}^P$ 
directly acting on the external fermion line yields a pinch
part.

Extracting the pinch parts of the diagrams represented by
fig.~\ref{fig:pin} and combining them with the usual \se\
we find a result of the form
\beqar
\lefteqn{\Sigma_T^{R_{\xi}, (2)}(p^2) + 
	 \Sigma_{T, Pin}^{(2)}(p^2) =} \no
\lefteqn{\quad = e^4 \Bigl( \tilde{f}(A_0, B_0) + 
		\tilde{g}(A_0, B_0; \xi_W) +
                \tilde{F}(T) \Big) .}
\label{eq:pinirr}
\eeqar
Here $\Sigma_T^{R_{\xi}, (2)}(p^2)$ denotes the usual \se\ in the
$R_{\xi}$-gauge as given in~(\ref{eq:zW}),
$\Sigma_{T, Pin}^{(2)}(p^2)$ are the pinch parts of the
one-particle irreducible two-loop triangle and box graphs, 
$\tilde{f}(A_0, B_0)$ and $\tilde{F}(T)$ are gauge independent
functions of scalar one-loop and two-loop integrals,
respectively, whereas $\tilde{g}(A_0, B_0; \xi_W)$ still depends
on the gauge parameter. This means that only the gauge parameter
dependence of the terms containing the scalar two-loop integrals
has disappeared, while the contributions involving the one-loop
integrals remain gauge dependent. 
This feature was in fact anticipated from the
structure of the result~(\ref{eq:zW}). Since the pinch parts
extracted from the two-loop triangle and box graphs are by
construction proportional to $(p^2 - m_Z^2)$, they cannot cancel
the gauge dependence still present in~(\ref{eq:zW}) for
$p^2 = m_Z^2$.

The complete two-loop contribution to the gauge boson
propagators in the PT is given as
\beqar
\lefteqn{\Delta_T^{PT, (2)}(p^2) = - i \Bigl\{
         \frac{\Sigma_T^{R_{\xi}, (2)}(p^2) +
         \Sigma_{T, Pin}^{(2)}(p^2)}{(p^2 - m^2)^2}} \no
&&  \qquad \quad - \frac{\Big(\Sigma_{T}^{R_{\xi}, (1)} (p^2) + 
     \tilde{\Sigma}_{T, Pin}^{(1)} (p^2)\Big)^2}{(p^2 - m^2)^3}
     \Bigr\} ,
\label{eq:pinprop}
\eeqar
where the first line contains the one-particle irreducible
contributions of~(\ref{eq:pinirr}) and the second line the reducible
ones. For the $Z$ propagator one also has to take into account
the mixing with the photon. 
Since $\tilde{\Sigma}_{T, Pin}^{(1)} (p^2)$ originates only
from reducible graphs, it is not the full one-loop pinch part.
In particular, it contains no contributions from one-loop
box diagrams.
However, these are necessary to cancel the gauge parameter
dependence of the one-loop gauge boson \se,
$\Sigma_{T}^{R_{\xi}, (1)} (p^2)$~\cite{pinch}. The reducible
contribution in~(\ref{eq:pinprop}) is therefore also
gauge parameter dependent and does not correspond to the
one-loop \se\ in the PT, $\Sigma_{T}^{PT, (1)}$.

To identify the quantity which has to be regarded as the
two-loop \se\ in the PT framework, we write the
propagator~(\ref{eq:pinprop}) as
\begin{displaymath}
\Delta_T^{PT, (2)}(p^2) = - i \Bigl\{
\frac{\Sigma_{T}^{PT, (2)}}{(p^2 - m^2)^2}
- \frac{(\Sigma_{T}^{PT, (1)})^2}{(p^2 - m^2)^3} 
\Bigr\} .
\end{displaymath}
This yields
\beqar
\lefteqn{\Sigma_{T}^{PT,(2)}(p^2) = \Sigma_T^{R_{\xi}, (2)}(p^2) +
         \Sigma_{T, Pin}^{(2)}(p^2)} \no
&& - \frac{1}{p^2 - m^2}
   \Bigl[\Bigl( \Sigma_{T}^{R_{\xi}, (1)} (p^2) +
                \tilde{\Sigma}_{T, Pin}^{(1)} (p^2) \Bigr)^2 \no
&& \qquad \qquad \; \; \; - \Bigl( \Sigma_{T}^{PT, (1)}(p^2) \Bigr)^2
  \Bigr] .
\label{eq:zWPT}
\eeqar
The two-loop \se\ constructed via the PT therefore involves both
one-particle irreducible and reducible contributions.

We have evaluated this quantity for the ``charged current'' 
light fermion contributions to the $Z$ \se\
and found that it is in fact independent of the
gauge parameters. The result will be given in the next section.

\subsection{The background-field method}

The BFM~\cite{Ab81} is a technique
for quantizing gauge theories without losing explicit gauge
invariance. This is achieved by decomposing in the Lagrangian 
the usual gauge field into a background field $\Vhat$ and a
quantum field $V$. By adding a suitable non-linear gauge-fixing
term, an effective action $\Ga[\Vhat]$ is obtained which is
invariant under gauge transformations of the background fields.
While the background fields are associated with the tree
propagators, the quantum fields appear only inside loops. 
In $\Ga[\Vhat]$, only
for the quantum fields a gauge fixing is needed.
We specify it by the quantum gauge parameter $\xiQ$.



The BFM is valid for all vertex functions to all orders of
perturbation theory. Its approach is very different from the PT.
While the PT seeks to obtain suitable 
``Green functions'' by eliminating their dependence
on the parameters of the $R_{\xi}$-gauge, the BFM is based on
the gauge invariance inherent in $\Ga[\Vhat]$. The fact
that the vertex functions fulfill simple Ward identities and
possess other desirable features is a direct consequence of
gauge invariance in the BFM~\cite{bgf}, 
whereas the PT provides no explanation for these properties. 

The comparison of these methods at the two-loop level
clearly shows the power of the BFM
approach. In the PT an involved rearrangement was necessary between
different Green functions contributing to a specific two-loop
process and between reducible and irreducible contributions. 
Contrarily, the vertex functions of the BFM are 
uniquely given in terms of Feynman rules. One simply
has to calculate the ``charged current'' diagrams shown in 
fig.~\ref{fig:diagSM}b using the BFM Feynman rules. The result is
just an ordinary \se\ which is guaranteed to be process independent.
For the special case 
$\xi_Q = 1$ it reads
\beqar
\lefteqn{\Sigma_{T, \xi_Q=1}^{BFM, (2)}(p^2) =
         \Sigma_{T, \xi=1}^{R_{\xi}, (2)}(p^2) - e^4 18 c_W^2 C} \no
\lefteqn{ \; \; \; \times \Bigl[
          4 (1 - D) p^2 B_0(p^2;0,0) B_0(p^2; m_W^2, m_W^2)} \no
\lefteqn{ \quad \; \; + (p^2 - m_Z^2)
          \Bigl( \frac{(3 - D) p^2 - m_W^2}{p^2 m_W^2}
          T_{13'4'}(m_W^2) } \no
\lefteqn{ \qquad \qquad \quad
          + \frac{1}{p^2} T_{23'4'}(p^2;m_W^2)} \no
\lefteqn{ \qquad \qquad \quad
          - 2 (3 - 2 D) T_{123'4'}(p^2;m_W^2, m_W^2)} \no
\lefteqn{ \qquad \qquad \quad
          + [p^2 - 4 (2 - D) m_W^2] } \no
\lefteqn{ \qquad \qquad \quad \; \;
          \times T_{1123'4'}(p^2;m_W^2,m_W^2, m_W^2)
          \Bigr) \Bigr] ,}
\label{eq:zWBFM}
\eeqar
where  $\Sigma_{T, \xi=1}^{R_{\xi}, (2)}(p^2)$ is the 
conventional \se\ in the `t Hooft-Feynman gauge, i.e.~the result
given in~(\ref{eq:zW}) with $\xi_W = 1$.
The result~(\ref{eq:zWBFM}) exactly coincides with the PT
expression~(\ref{eq:zWPT}). Therefore we find here the same
connection between the PT and the BFM as in one-loop order, 
i.e.~the BFM contains the PT results as a special case.

We would like to stress that the case $\xi_Q=1$
displayed here is not distinguished in the BFM.
All properties derivable from the gauge invariant effective
action, e.g.~the simple BFM Ward identities, hold for all finite 
values of $\xi_Q$. As an example, we list a Ward identity valid
for the $Z$ \se\ in the BFM to all orders
\beq
p^{\mu} \Sigma^{\hat Z \hat Z}_{\mu \nu}(p) - i M_Z
\Sigma^{\hat\chi \hat Z}_{\nu}(p) = 0 ,
\eeq
where the conventions of~\cite{bgf} are used. We checked the
validity of this Ward identity at the two-loop level by explicit
computation for arbitrary $\xi_Q$. 
The Ward identity fulfilled by the conventional 
two-loop $Z$ \se\ was given in~\cite{Red}. It involves in
addition the \se\ of the unphysical scalar $\chi$ and reducible
contributions.

In analogy to the one-loop case, this means that the
special case singled out in the PT is not unique.
When studying applications of two-loop \ses, it has to be
investigated whether these are influenced by the ambiguity
corresponding to different choices of $\xi_Q$. 





\section{ANALYTIC APPROACHES AND HYPERGEOMETRIC FUNCTIONS}
\subsection{Introductory remarks}
As explained above, the algebraic reduction of two-loop \ses\
leads to four types of two-loop scalar integrals,
$T_{234}$, $T_{1234}$, $T_{12345}$ and $T_{134}$. The last 
integral follows from $T_{234}$ by setting $p^2 = 0$ (see below),
so one needs to study three scalar two-loop integrals.
For convenience, we will in the following relabel the integral 
$T_{234}$ as $T_{123}(p^2;m_1^2,m_2^2,m_3^2)$.

In this section we discuss analytic results in an arbitrary 
number of dimensions $D$ for these integrals. They will be in the
form of generalized hypergeometric functions, that is multiple 
series of ratios of $p^2$ and the masses. We also discuss the 
expansion of these results around $D=4$.
Only in those cases which fulfill the condition 
\begin{equation}
 p^2 m_1^2 m_2^2 m_3^2 \prod 
 \left(p^2-\left(m_1 \pm m_2 \pm m_3\right)^2\right)=0
\end{equation}
for the three-particle cuts, the finite part can be expressed in 
terms of polylogarithms. A typical example is given below:
\begin{eqnarray}
&&\hskip-7mm T_{123}(p^2;0,0,m^2) = -m^2 
 \left(\frac{m^2}{4\pi\mu^2}\right)^{- 2\delta}  
 \label{T00m} \nonumber \\
 && \hskip-7mm  \frac{\Gamma^2(1\!-\!\delta) 
 \Gamma(\delta) \Gamma(-1\!+\!2\delta)}
        {\Gamma(2\!-\!\delta)} 
     {}_2 F_1(\delta,2 \delta-1;2-\delta;z) \nonumber\\
&& \hskip-7mm   = m^2 \left[\frac{1}{2\delta^{2}} + \frac{1}{\delta} 
               \left(\frac{3}{2}
               -L_m -\frac{z}{4} \right) +3 + \frac{3}{2}\zeta(2)  \right.
               \nonumber\\
&& \hskip+5mm  +L_m^2-\left(3-\frac{z}{2}\right)L_m -\frac{13}{8} z + Li_2(z)
     \nonumber\\
&& \hskip+5mm  \left.- \frac{1}{2}\left(\frac{1}{z}-z \right)\log(1-z)\right] 
\end{eqnarray}
where
$z=p^2/m^2$,
$L_m = \gamma + \ln (m^2/4\pi\mu^2)$, $\gamma$ is the Euler constant 
and $\zeta(2)=\pi^{2}/6$.
 Another example is the massive vacuum diagram 
$T_{123}(0;m_1^2,m_2^2,m_3^2)$ which in $D$ dimensions is given 
in terms of four Appell $F_4$ functions 
but in four dimensions contains only dilogarithms~\cite{Tausk}.
In these specific cases one works with Feynman parameters or dispersion 
relations \cite{Scharf}.
In the general mass case it turns out that $x$-space techniques and in 
particular the Mellin-Barnes representations are also useful. 
In the following subsections we will give an example of the 
application of dispersion relations and of the Mellin-Barnes representation.
\boldmath
\subsection{The integral $T_{123}$}
\unboldmath
Let us first start with the rederivation of the so-called London 
transport diagram $T_{123}(p^2;m_1^2,m_2^2,m_3^2)$ by means of 
dispersion relations. 
The imaginary part or the discontinuity $\Delta T_{123}$ is given by
\begin{eqnarray}
 && \hskip-7mm  ImT_{123}(p^{2};m_1^2,m_2^2,m_3^2)  = \label{imag} \\
 && \hskip-7mm  
\frac{\Delta T_{123}(p^{2};m_{1}^{2},m_{2}^{2},m_{3}^{2})}{2 i} =
 \nonumber\\
 && \hskip-7mm  - \pi  (4\pi\mu^2)^{2 \delta}
            \frac{\Gamma^2(1\!-\!\delta)}{\Gamma^2(2\!-\!2 \delta)}
          \Theta\left(p^2\!-\!(m_1\!+\!m_2\!+\!m_3)^2\right)\nonumber\\
 && \hskip-7mm  \int\limits _{(m_2 + m_3)^2 } ^{(\sqrt
          p^2-m_1)^2}
        \!\!\! {\mbox{d} s}
        \frac{\lambda^{\frac{1}{2}-\delta} (s,m_2^2,m_3^2)}{s^{1-\delta}}
        \frac{\lambda^{\frac{1}{2}-\delta} (p^2,s,m_1^2)}{{(p^2)^{1-\delta}}}
              \nonumber \\
 && \hskip-7mm  = \frac{1}{4 \pi}\!\!\! 
\int\limits _{(m_2 + m_3)^2 } ^{(\sqrt p^2-m_1)^2}
      \!\!\! {\mbox{d} s}
      \Delta B_0(s;m_2^2,m_3^2) \Delta B_0(p^2;s,m_1^2) \nonumber
\end{eqnarray}
where $\lambda(a,b,c)=(a-b-c)^2-4 b c$ is the 
K\"{a}ll\'{e}n function.
The dispersion relation reads
\begin{eqnarray}
 &&\hskip -7mm  T_{123}(p^{2};m_{1}^{2},m_{2}^{2},m_{3}^{2}) = \label{disp} 
    \nonumber \\
 && \hskip -7mm \frac{1}{2\pi i}\!\!\!
 \!\!\! \int\limits_{(m_{1}+m_{2}+m_{3})^2}^{\infty}\!\!\!\!\!\!\!\! 
\mbox{d} z \frac{1}{z-p^2}
     \Delta T_{123}(z;m_{1}^{2},m_{2}^{2},m_{3}^{2})\nonumber\\
 && \hskip -7mm  = \frac{1}{4 \pi^2}
\int\limits_{(m_{2}+m_{3})^2}^{\infty}\!\!\!
\mbox{d} s
      \,\,\Delta B_0(s;m_2^2,m_3^2) \nonumber\\
 && \hskip -7mm \times  
\int\limits_{(\sqrt s+m_1)^2}^{\infty}\frac{\mbox{d} z}{z-p^2}\,\,
      \Delta B_0(z;s,m_1^2) .
\end{eqnarray}
Using the expansion
\begin{equation}
 \frac{1}{z-p^2}=\frac{1}{z} \sum_{k=0}^{\infty} \left(\frac{p^2}{z}\right)^k
\end{equation}
we perform first the integration over $z$:
\begin{eqnarray}
 && \hskip-7mm A = \sum_{k=0}^{\infty} (p^2)^k \int\limits_{u}^{\infty}
                   \frac{ \mbox{d} z}{z^{k+2-\delta}}
        (z-u)^{1/2-\delta} (z-v)^{1/2-\delta}\label{A}\nonumber\\
 &&    =\sum_{k=0}^{\infty} (p^2)^k u^{-\delta-k}
     B(k+\delta,3/2-\delta)
     \nonumber\\
  && \times {}_2 F_1(\delta-1/2,k+\delta;k+3/2;v/u)
\end{eqnarray}
where $u\,=\,(m_1+\sqrt s)^2$ and $v\,=\,(m_1-\sqrt s)^2$. 
One can transform the Gauss hypergeometric function ${}_2 F_1$ 
using relations which one can find in
\cite{Prudnikov}
leading to $T_{123}(p^{2};m_{1}^{2},m_{2}^{2},m_{3}^{2})$
\begin{eqnarray}
  && \hskip-7mm T_{123}(p^{2};m_{1}^{2},m_{2}^{2},m_{3}^{2}) =         
             -\left(4\pi\mu^2 \right)^{\delta} \label{ints}\nonumber \\
  && \hskip-7mm \times  
    \frac{\Gamma(1\!-\!\delta) \Gamma(\delta\!-\!1)}{\Gamma(2\!-\!2\delta)}
     \!\!\!
     \int\limits_{(m_{2}+m_{3})^2}^{\infty} \!\!\!\!\mbox{d} s\, s^{\delta-1}
    \lambda^{\frac{1}{2}-\delta} (s,m_2^2,m_3^2)\nonumber\\
  && \hskip-7mm 
    \left[\frac{m_1^2}{s} \left(\frac{m_1^2}{4\pi\mu^2}\right)^{-\delta}
    \!\!
    F_4(1,2\!-\!\delta;2\!-\!\delta,2\!-\!\delta;\frac{p^2}{s},\frac{m_1^2}{s})
    \right. \nonumber\\
  &&\left.-\left(\frac{s}{4\pi\mu^2}\right)^{-\delta}
     \!\!
     F_4(1,\delta;2\!-\!\delta,\delta;\frac{p^2}{s},\frac{m_1^2}{s})\right].
\end{eqnarray}
Since in (\ref{disp}) the $z$ integration up to some factors 
represents the one-loop self-energy, we find as a byproduct
\begin{eqnarray}
\label{oneloop}
 && \hskip-7mm
 B_0(p^2;m_1^2,m_2^2) =\Gamma(\delta-1) \nonumber \\
 &&\hskip-7mm \times 
 \left[\frac{m_1^2}{m_2^2} \left(\frac{m_1^2}{4\pi\mu^2}\right)^{-\delta}
    \!\!\!\!\!
    F_4(1,2\!-\!\delta;2\!-\!\delta,2\!-\!\delta; 
\frac{p^2}{m_2^2},\frac{m_1^2}{m_2^2})
    \right.  \nonumber \\
  &&\left.-\left(\frac{m_2^2}{4\pi\mu^2}\right)^{-\delta}
     \!\!\!\!\!
     F_4(1,\delta;2\!-\!\delta,\delta;\frac{p^2}{m_2^2},\frac{m_1^2}{m_2^2})
     \right].
\end{eqnarray}
 
Using the definition of the $F_4$ functions we can easily perform 
the integration over $s$. 
After some manipulations  we obtain the result for the 
London transport diagram 
which agrees with the one derived using $x$-space technique and 
Mellin-Barnes representation \cite{Buza}. 
The result is presented in table (\ref{tab:small}),
\begin{table*}[htb]
\setlength{\tabcolsep}{1.5pc}
\newlength{\digitwidth} \settowidth{\digitwidth}{\rm 0}
\catcode`?=\active \def?{\kern\digitwidth}
\caption{The small $p^2$ result}
\label{tab:small}
\begin{tabular*}{\textwidth}{@{}l@{\extracolsep{\fill}}}
\hline
 $T_{123}(p^{2};m_{1}^{2},m_{2}^{2},m_{3}^{2}) =
-m_{3}^{2} \left(\frac{m_{3}^{2}}{4\pi\mu^{2}}\right)^{2(\nu-1)}
\times
$    \\
$\left\{   
z_{1}^{\nu}z_{2}^{\nu}\;
\Gamma^2(-\nu)\;
F_{C}^{(3)}(1,1+\nu; 1+\nu,1+\nu,1+\nu;z_{1},z_{2},z_{3})
\right.$   \\
$-z_{1}^{\nu}\;
\Gamma^2(-\nu)\;
F_{C}^{(3)}(1,1-\nu; 1+\nu,1-\nu,1+\nu;z_{1},z_{2},z_{3})
$     \\
$-z_{2}^{\nu}\;
\Gamma^2(-\nu)\;
F_{C}^{(3)}(1,1-\nu; 1-\nu,1+  \nu,1+\nu; z_{1},z_{2},z_{3})
$   \\
$\left.
-\Gamma(\nu)\Gamma(-\nu)\Gamma(1-2\nu)\;
F_{C}^{(3)}(1-2\nu,1-\nu; 1-\nu,1-\nu,1+\nu; z_{1},z_{2},z_{3})
\right\}$          \\
\hline
\multicolumn{1}{@{}p{120mm}}{}
\end{tabular*}
\end{table*}
 where $z_{i}=m_{i}^{2}/m_{3}^{2}\,,\,i=1,2\,,z_{3}=p^{2}/m_{3}^{2}$ 
and $\nu=1-\delta$.

In the result we recognize a special instance of the
Lauricella functions \cite{Exton}
defined by
\begin{eqnarray}
&& \hskip-7mm  F_{C}^{(n)}(a,b;c_1,\ldots,c_n;z_1,\ldots,z_n) = \\
&& \hskip-7mm \sum_{k_1,\ldots,k_n=0}^{\infty}
  \frac{(a)_{k_1+\ldots+k_n} (b)_{k_1+\ldots+k_n}}{(c_1)_{k_1}\,\cdots\,
  (c_n)_{k_n}}
 \frac{z_1^{k_1}\,\cdots\,z_n^{k_n}}{k_1!\,\cdots\,k_n!} \;\;, \nonumber
\end{eqnarray}
 where $(a)_k=\Gamma(a+k)/\Gamma(a)$.
The defining multiple series converges for
\begin{equation}
\sqrt{|z_{1}|}+\ldots+\sqrt{|z_{n}|}\,<\,1\;.
\label{Lconv}
\end{equation}
The individual series above converge for
$m_{1}+m_{2}+\sqrt{|p^{2}|}<m_{3}$.
Collecting powers in $p^2$, however, the total sum converges
due to analyticity up to the next singularity on the physical sheet
given by the threshold condition
\begin{equation}
 |p^{2}|\,<\,(m_{1}+m_{2}+m_{3})^2\,,
\end{equation}
provided that the coefficients themselves do exist, which is the case for
\begin{equation}
m_{1}+m_{2}\,<\,m_{3}\,.
\label{converge}
\end{equation}

A Lauricella function in the arguments
$z_i$ for $i=1,\ldots,n$ can be
analytically continued to a sum of two Lauricella functions
in the arguments
$x_{i}=z_{i}/z_{n}$ for $i=1,\ldots,n-1$ and $x_{n}=1/z_{n}$
by the following relation \cite{Exton}
\begin{eqnarray}
&& \hskip-7mm
F_C^{(n)}(a,b;c_1,\ldots,c_n;z_1,\ldots,z_n) =
\\
&& \hskip-7mm
\phantom{+}
f_1 F_C^{(n)}(a\!,\!1\!+\!a\!-\!c_n;c_1,\ldots,c_{n-1},\!1\!-\!b\!+\!a;
x_{1,\ldots,n})
\nonumber\\
&&
\hskip-7mm
+
f_2 F_C^{(n)}(b\!,\!1\!+\!b\!-\!c_n;c_1,\ldots,c_{n-1},\!1\!-\!a\!+\!b;
x_{1,\ldots,n})
\;.\nonumber
\end{eqnarray}
where
\begin{eqnarray}
&& f_1=\frac{\Gamma(c_n)\Gamma(b-a)}{\Gamma(b)\Gamma(c_n-a)}(-z_n)^{-a},\,\,\,\,
 \nonumber \\
&& f_2=\frac{\Gamma(c_n)\Gamma(a-b)}{\Gamma(a)\Gamma(c_{n}-b)}(-z_n)^{-b}.
\end{eqnarray}
Applied to tab. (\ref{tab:small}) this yields,
as one coefficient vanishes,
a total of seven transformed Lauricella functions, which is 
also given in the tab. (\ref{tab:large}).
\begin{table*}[hbt]
\setlength{\tabcolsep}{1.5pc}
\catcode`?=\active \def?{\kern\digitwidth}
\caption{The large $p^2$ result}
\label{tab:large}
\begin{tabular*}{\textwidth}{@{}l@{\extracolsep{\fill}}}
\hline
$ T_{123}(p^{2};m_{1}^{2},m_{2}^{2},m_{3}^{2}) =
  -\left(\frac{-p^{2}}{4\pi\mu^{2}}\right)^{2\nu-2}(-p^{2}) \times
  $ \\
$ \left\{(-x_{1})^{\nu}(-x_{3})^{\nu}\Gamma^2(-\nu)
  \;F_{C}^{(3)}(1,1-\nu;1+\nu,1-\nu,1+\nu;x_{1},x_{2},x_{3})\right.
  $ \\
$ +(-x_{2})^{\nu}(-x_{3})^{\nu}\Gamma^2(-\nu)
  \;F_{C}^{(3)}(1,1-\nu;1-\nu,1+\nu,1+\nu;x_{1},x_{2},x_{3})
  $ \\
$ +(-x_{1})^{\nu}(-x_{2})^{\nu}\Gamma^2(-\nu)
  \;F_{C}^{(3)}(1,1-\nu;1+\nu,1+\nu,1-\nu;x_{1},x_{2},x_{3})
  $ \\
$ +(-x_{1})^{\nu}\frac{\Gamma^2(\nu)\Gamma(-\nu)\Gamma(1-\nu)
  }{\Gamma(2\nu)}
\;F_{C}^{(3)}(1-\nu,1-2\nu;1+\nu,1-\nu,1-\nu;x_{1},x_{2},x_{3})
  $ \\
$ +(-x_{2})^{\nu}\frac{\Gamma^2(\nu)\Gamma(-\nu)\Gamma(1-\nu)
  }{\Gamma(2\nu)}
\;F_{C}^{(3)}(1-\nu,1-2\nu;1-\nu,1+\nu,1-\nu;x_{1},x_{2},x_{3})
  $ \\
$ +(-x_{3})^{\nu}\frac{\Gamma^2(\nu)\Gamma(-\nu)\Gamma(1-\nu)
  }{\Gamma(2\nu)}
\;F_{C}^{(3)}(1-\nu,1-2\nu;1-\nu,1-\nu,1+\nu;x_{1},x_{2},x_{3})
  $ \\
$ +\left. \frac{\Gamma^3(\nu)\Gamma(1-2\nu)}{\Gamma(3\nu)}
\; F_{C}^{(3)}(1-3\nu,1-2\nu;1-\nu,1-\nu,1-\nu;
x_{1},x_{2},x_{3})\right\}\,.
  $ \\
\hline
\multicolumn{1}{@{}p{120mm}}{}
\end{tabular*}
\end{table*}
Now, this expression is valid for
\begin{equation}
 |p^{2}|\,>\,(m_{1}+m_{2}+m_{3})^{2}\,.
\end{equation}

One may wonder what the relation is between the large $p^{2}$ expansion
of tab. (\ref{tab:large}) and that given in \cite{Smirnov}. In the latter
approach the various terms in the $p^{2}$ expansion are obtained
from the expansion of subgraphs. The subgraphs are obtained by
distributing the momentum $p$ over the propagators in all possible ways.
In the case of the London transport diagram one has the
following subgraphs: the diagram itself, the three diagrams where
one internal line is removed and the three diagrams where two
internal lines have been removed.
 
Following the analysis of \cite{Smirnov} one can easily find the
first term of each of the contributing
series. For the subgraph representing the whole diagram the first term in
the series should be the massless diagram. This series then
corresponds to the last term in tab. (\ref{tab:large}).
The series which originates from the subgraph where two lines
have been removed, e.g. 1 and 2, starts
with the product of two massive tadpoles. They contribute a
factor $\left(m_{1}^{2}m_{2}^{2}\right)^{\nu}$
which can be identified with the third term in tab. (\ref{tab:large}).
The remaining subgraphs
are obtained by removing one internal line, e.g. line 3.
This yields a series starting with a massive
tadpole proportional to $(m_{3}^{2})^{\nu}$.
This is the sixth term in tab. (\ref{tab:large}).
Thus the seven series in tab. (\ref{tab:large}) can be related
directly to the seven subgraphs
which are required for the method of \cite{Smirnov}.

For the special case $p^2=0$, $F_{C}^{(3)}$ reduces to the 
Appell function $F_4$
 and we recover the result in the literature \cite{Tausk} 
for the vacuum diagram. 
On the other hand taking $m_1=m_2=0$ only one of the Lauricella 
functions remains and becomes a Gauss hypergeometric 
function ${}_2 F_1$ giving the result
\begin{eqnarray}
&& \hskip-7mm 
   T_{123}(p^{2};0,0,m^2) = m^2 
       \left(\frac{m^2}{4\pi\mu^2}\right)^{-2 \delta}
\\
&& \hskip-7mm
 \Gamma(1\!-\!\delta) \Gamma(\delta\!-\!1)\Gamma(2 \delta\!-\!1)
   {}_2 F_1 (2 \delta\!-\!1,\delta;2\!-\!\delta;z)\,, \nonumber
\end{eqnarray}
which is the same as (\ref{T00m}).

In the following the general $D$ dimensional expression for the
2-loop London transport diagram will be expanded
in $\delta\,=\,(4-D)/2\,=\,1-\nu$ for small $|p^{2}|$. 

The following combination \cite{Berends}
of the general massive case with massless cases
is chosen in such a way that the infinite parts cancel
\begin{eqnarray}
&& \hskip-7mm T_{123N}(p^{2};m_{1}^{2},m_{2}^{2},m_{3}^{2}) = 
\label{DefT123N}                 \nonumber \\
&& \hskip-7mm T_{123}(p^{2};m_{1}^{2},m_{2}^{2},m_{3}^{2})
- T_{123}(p^{2};m_{1}^{2},0,m_{3}^{2}) \nonumber\\
&& \hskip-7mm
-T_{123}(p^{2};0,m_{2}^{2},m_{3}^{2})
+T_{123}(p^{2};0,0,m_{3}^{2})
\,.
\end{eqnarray}
It is this combination which will be calculated in two
independent ways. 

An analytic form is obtained by expansion of
the Lauricella functions and their coefficients
in $\delta$, where the first and the second
logarithmic derivatives of the $\Gamma$-function occur at integer arguments
\begin{eqnarray}
&& \hskip-7mm  \psi(n+1)=-\gamma+\sum_{k=1}^{n}\frac{1}{k}, \nonumber\\
&& \hskip-7mm  \psi'(n+1)=\zeta(2)-\sum_{k=1}^{n}\frac{1}{k^{2}}. 
\end{eqnarray}
 
The $1/\delta^2$ and $1/\delta$ terms indeed drop out
in the result
and a finite combination of various multiple series remains.
A good check is provided by the cancellation of $\gamma$
in this finite expression.\\
For small $|p^{2}|$, i.e. the region $|p^{2}|\,<\,m_{3}^{2}$,
one finds
\begin{eqnarray}
&&\hskip-7mm
T_{123N}(p^{2};m_{1}^{2},m_{2}^{2},m_{3}^{2})/m_{3}^{2}=
\label{smallfin} \nonumber \\
&&\hskip-7mm
-\!\!\!\sum_{m,n=1,k=0}^{\infty} \!\!
\frac{(t-2)!(t-1)!}
     {(m-1)!m!(n-1)!n!(k+1)!k!}z_{1}^{m}z_{2}^{n}z_{3}^{k}
\nonumber\\
&&\hskip-7mm
\Big[
\big\{\psi(t)+\psi(t\!-\!1)-\psi(m)-\psi(m\!+\!1)+\log(z_{1})\big\}
\nonumber\\
&&\hskip-5mm
\times
\big\{\psi(t)+\psi(t\!-\!1)-\psi(n)-\psi(n\!+\!1)+\log(z_{2})\big\}
\nonumber\\
&&\hskip-5mm
+\psi'(t)+\psi'(t\!-\!1)
\Big] \,
\end{eqnarray}
with $t=m+n+k$.
\boldmath
\subsection{The integral $T_{1234}$}
\unboldmath
With the dispersion method described above we derive the {\sl small} $p^2$ {\sl
result} for $T_{1234}$ in $D$ dimensions.
The discontinuity $\Delta T_{1234}$ is a sum of a two and a three-particle cut,
which we denote by $\Delta T_{1234}^{(2)}$ and 
$\Delta T_{1234}^{(3)}$, respectively. The two-particle cut is given by
\begin{eqnarray}
&&
  \Delta T_{1234}^{(2)}(p^2;m_1^2,m_2^2,m_3^2,m_4^2) \nonumber \\
&& = \Delta B_0(p^2;m_1^2,m_2^2) B_0(m_1^2;m_3^2,m_4^2)\, . \label{DeltaT12342}
\end{eqnarray}
Inserting this result in the dispersion integral gives
\begin{eqnarray}
 T_{1234}^{(2)}(p^2) &=& \frac{1}{2\pi i} \!\!\!
                 \int\limits_{(m_1+m_2)^2}^{\infty} \!\!\! \mbox{d} z\, 
\frac{1}{z-p^2} \Delta T_{1234}^{(2)}(z) \nonumber\\
  &=&B_0(p^2;m_1^2,m_2^2) B_0(m_1^2;m_3^2,m_4^2)\nonumber.
\end{eqnarray}
The discontinuity $\Delta T_{1234}^{(3)}$ is related to the three 
particle-cut given in (\ref{imag}) but with an additional propagator 
and therefore we
get the dispersion relation
\begin{eqnarray}
  && \hskip-7mm T_{1234}^{(3)}(p^2;m_i^2) =
    -(4\pi\mu^2)^{2 \delta}
      \frac{\Gamma^2(1-\delta)}{\Gamma^2(2-2 \delta)}
       \label{T3disp}\\
  &&\hskip-7mm
 \times\!\!\!\! \int\limits_{(m_{3}+m_{4})^2}^{\infty}
\!\!\!\!\!\!\!\!\!\! \mbox{d} s\,\,
\frac{\lambda^{\frac{1}{2}-\delta} (s,m_3^2,m_4^2)}
       {s^{1-\delta}(s-m_1^2-i\epsilon)}
\!\!\!\!\int\limits_{(\sqrt s+m_2)^2}^{\infty}\!\!\!\!\!\!\!\!\!\! \mbox{d} z
      \frac{\lambda^{\frac{1}{2}-\delta} (s,z,m_2^2)}{z^{1-\delta} (z-p^2)}
      \nonumber.
\end{eqnarray}
After performing the $z$ integration we obtain:
\begin{eqnarray}
&& \hskip-7mm T_{1234}^{(3)}(p^2;m_i^2) =
-(4\pi\mu^2)^{\delta}
 \frac{\Gamma(1-\delta)\Gamma(\delta-1)}{\Gamma(2-2 \delta)}\label{T3s} 
  \nonumber \\
&& \hskip-7mm  \times \!\!\!
\int\limits_{(m_{3}+m_{4})^2}^{\infty} \!\!\! \mbox{d} s\, 
\frac{s^{\delta-1}}{s-m_1^2} \lambda^{\frac{1}{2}-\delta} (s,m_3^2,m_4^2) 
\nonumber\\
&& \hskip-7mm
\times \!
    \left[\frac{m_2^2}{s}\! \left(\frac{m_2^2}{4\pi\mu^2}\right)^{-\delta}
    \!\!\!\!
    F_4(1,2\!-\!\delta;2\!-\!\delta,2\!-\!\delta;\frac{p^2}{s},\frac{m_2^2}{s})
    \right. \nonumber\\
&& 
  \left.-\left(\frac{s}{4\pi\mu^2}\right)^{-\delta}\!
     \!\!\!
     F_4(1,\delta;2\!-\!\delta,\delta;\frac{p^2}{s},\frac{m_2^2}{s})\right].
\end{eqnarray}

Expanding $(s-m_1^2)^{-1}$ in $m_1^2/s$ and performing the integration  
over $s$ one gets
\begin{eqnarray}
 && \hskip-7mm T_{1234}^{(3)}(p^2;m_i^2) = \frac{\Gamma(1-\delta) 
                       \Gamma(1+\delta)}{\delta}
         \left(\frac{m_4^2}{4\pi\mu^2}\right)^{- 2\delta}
                   \label{T3} \nonumber \\
 && \hskip-7mm 
\times \sum_{m,n,k,l=0}^{\infty} z_1^k z_2^n (1-z_3)^l z_4^m
      \nonumber\\
 && \hskip-7mm 
\times \frac{\Gamma(1+m+n)\Gamma(1+\delta+m+n+k+l)}{\Gamma(2-\delta+m)
      m!n!l!}\nonumber\\
 && \hskip-7mm
\times \left(z_2^{1-\delta}
\frac{\Gamma(2-\delta+m+n)\Gamma(2+m+n+k+l)}
     {\Gamma(2-\delta+n)\Gamma(2(m+n+k)+4+l)}\right.\nonumber\\
 && \hskip-4mm
    \left.-\frac{\Gamma(\delta+m+n)\Gamma(2 \delta+m+n+k+l)}
           {\Gamma(\delta+n)\Gamma(2(m+n+k)+2+2 \delta+l)}
           \right).
\end{eqnarray}
where $z_i\,=\,m_i^2/m_4^2$ with $i=1,2,3$ and $z_4\,=\,p^2/m_4^2$.
Note that the contribution from the three-particle cut is written in 
terms of multiple series which are not Lauricella functions 
anymore but they belong to a special class of generalized 
hypergeometric functions.
To our knowledge they have not been studied in the mathematical literature.
 
With these results $T_{1234}$ becomes
\begin{eqnarray}
 && T_{1234}(p^{2};m_i^2) = T_{1234}^{(3)}(p^2;m_i^2) \label{T1234} \nonumber \\
  && + B_0(p^2;m_1^2,m_2^2) B_0(m_1^2;m_3^2,m_4^2).
\end{eqnarray}

Since the analytic continuation of the above generalized 
hypergeometric functions is not known to us we use the 
Mellin-Barnes representation technique to derive the 
{\sl large} $p^2$ expansion. In this method a massive 
propagator is represented in the following way:
\begin{eqnarray}
  \frac{1}{(k^{2}-m^{2})^{\alpha}} &=&
\frac{1}{\Gamma(\alpha)}\,
\frac{1}{2\pi i}
\int\limits_{-i\infty}^{+i\infty}ds\,
\frac{(-m^{2})^{s}}{(k^{2})^{\alpha+s}} \label{Mellin} \nonumber \\
&& \times \Gamma(-s)\Gamma(\alpha+s)\, ,
\end{eqnarray}
where the integration contour in the $s$ plane must separate
the series of poles of $\Gamma(-s)$ on the right
from the series of poles of $\Gamma(s+\alpha)$ on the left. 
In the expression for $T_{1234}$ we
apply (\ref{Mellin}) with $\alpha=1$
to all propagators, thereby relating the general massive case
to the massless one, but with the
propagators raised to arbitrary powers 
$1\!+\!s_{1},1\!+\!s_{2},1\!+\!s_{3}$,$1\!+\!s_4$.
The corresponding result is well-known, see e.g.\cite{Tausk}.
Closing the integration contours in a way that the convergence 
is guaranteed we get seven quartic series each of which 
can again be identified with the subgraph analysis for 
large $p^2$ expansions in \cite{Smirnov}.

The general small $p^2$ result for $T_{1234}$ can be expanded
in $\delta$.
The following combination of the general massive case with a 
massless case is chosen in such a way that the infinite 
parts cancel \cite{Berends}
\begin{eqnarray}
\label{finite}
 &&\hskip-7mm  T_{1234N}(p^2;m_1^2,m_2^2,m_3^2,m_4^2)  = \label{TN} \\ 
 &&\hskip-7mm   T_{1234}(p^2;m_1^2,m_2^2,m_3^2,m_4^2) \!
  -\! T_{1234}(p^2;m_1^2,m_2^2,0,0)\nonumber.
\end{eqnarray}
The result can be found in \cite{Bauber}.

As an example of a special case of these general results we 
choose $m_2=m_3=0$ and $m_1=m_4=m$. Eq. (\ref{T3}) gives
\begin{eqnarray}
  T_{1234}^{(3)}(p^2) &=& \!\!\!\! -\frac{\Gamma(\delta) \Gamma(2 \delta) 
             \Gamma(1-\delta)}
       {(1-\delta)(1-2 \delta)}\left(\frac{m^2}{4\pi\mu^2}\right)^{-2 \delta} 
        \nonumber \\
 &&  \times  {}_2 F_1(2 \delta,\delta;2-\delta;x)\,.
\end{eqnarray}
To this the contribution from the two-particle cut 
$B_0(p^2;m^2,0)B_0(m^2;0,m^2)$ should be added
\begin{eqnarray}
   T_{1234}^{(2)}(p^2) &=&
 \frac{\Gamma^2(1+\delta)}{\delta^2 (1-\delta) (1-2 \delta)}
 \left(\frac{m^2}{4\pi\mu^2}\right)^{-2 \delta} \nonumber  \\
 &&  \times  {}_2 F_1(1,\delta;2-\delta;x) .
\end{eqnarray}
Expanding in $\delta$ we get
\begin{eqnarray}
 && \hskip-7mm  T_{1234}(p^2;m^2,0,0,m^2) =  \nonumber\\
 && \hskip-7mm \frac{1}{2 \delta^{2}}+ \frac{1}{2 \delta}
 \left\{5 - 2 L_m+ 2 \left(\frac{1-x}{x}\right)
 \ln \left(1 - x\right)\right\}\nonumber \\
 && \hskip-7mm 
 + \frac{19}{2} - \frac{1}{2} \zeta(2) + L_m^2 - 5 L_m
 \nonumber\\
 && \hskip-7mm
 -2 L_m \!\left(\frac{1-x}{x}\right)\! \ln \left(1 - x\right)+ 
5 \left(\frac{1-x}{x}\right) \ln \left(1 - x\right)
 \nonumber\\
 && \hskip-7mm 
 - \left(\frac{1-x}{x}\right) \ln^2 \left(1 - x\right)  
                         - \frac{1}{x} Li_2(x) .
\end{eqnarray}

We conclude this section with some remarks on the master diagram 
$T_{12345}(p^2)$ in arbitrary number of dimensions. 
This case turns out to pose problems. For instance when 
one wants to apply the Mellin-Barnes representation, one needs 
the massless master diagram with arbitrary powers for the propagators. 
This expression is not (yet) available in the literature, 
which is related to the fact that the structure of the master 
diagram is not anymore of a self-energy insertion type.

\section{ANALYTIC APPROACHES AND ELLIPTIC INTEGRALS}

In this section we inspect the imaginary parts of
the London transport diagram $T_{123}$ and of $T_{1234}$.
It turns out that they can be calculated in four dimensions
in terms of complete elliptic integrals. These are well known
functions and thus the results are of analytic interest.
Furthermore fast and precise algorithms for the calculation
of the elliptic integrals are available. Therefore the results provide
also an efficient way to calculate the imaginary parts numerically.

\subsection{The imaginary part of the London transport diagram}

As can be seen from (\ref{imag}) the imaginary part of
$T_{123}$ is finite in four dimensions and reads with a
factorization of the K\"{a}ll\'{e}n functions
\begin{eqnarray}
 &&\!\! Im(T_{123}(p^2;m_i^2))= 
      \frac{1}{2 i} \Delta T_{123}(p^2,m_i^2) \nonumber \\
 &&  =
     -\frac{\pi}{p^2}
          \int\limits _{x_2} ^{x_3} { dt \over t } 
          \sqrt { (t-x_1) (t-x_2) } \nonumber \\
 &&  \;\;\;\;\;\;\;\;\;\;\;\;\;\; \times \sqrt{(x_3-t) (x_4-t) },
          \label{ImT123short}
\end{eqnarray}
with
\begin{eqnarray*}
 && x_1=(m_1-m_2)^2, \; x_2=(m_1+m_2)^2, \nonumber \\
 &&  x_3=(p-m_3)^2, \;  x_4=(p+m_3)^2, \nonumber \\  
 && x_1\le x_2 \le x_3 \le x_4. \nonumber
\end{eqnarray*}
The integration limits are zeros of the square roots, and thus
(\ref{ImT123short}) leads to complete elliptic integrals, defined
by
\begin{eqnarray}
{\rm K}(x) &=& {\int\limits _0 ^1 {dt \over \sqrt{(1-t^2)(1- x ^2 t^2)}}}
  \nonumber \\ 
  &=& {\pi \over 2}{_2 F _1}(-{1 \over 2},{1 \over 2},1,x^2), \\
{\rm E}(x) &=& {\int\limits _0 ^1 {dt {(1- x^2 t^2)}
               \over \sqrt{(1-t^2)(1- x ^2 t^2)} }} \nonumber \\
  &=& {\pi \over 2}{_2 F _1}({1 \over 2},{1 \over 2},1,x^2), \\
 \Pi(c,x) &=& {\int\limits _0 ^1 {dt  \over {(1-c t^2)
           \sqrt{(1-t^2)(1- x ^2 t^2)} }}} \nonumber \\
  &=& {\pi \over 2} F_1 ({1 \over 2};1,{1 \over 2};1;c,x^2),
\end{eqnarray}
with the Gauss hypergeometric function ${_2 F_1}$ and the Appell function
$F_1$ \cite{Exton,Appell,Srivastava}.
Reduction of (\ref{ImT123short}) to the Legendre normal form
of the elliptic integrals \cite{Erd,Abram}
by decomposition into partial fractions and partial integration
yields after some algebra
\begin{eqnarray}
 &&\!\!\!\!\! \!\!\!\!
 Im \left( T_{123}(p^2;m_1^2,m_2^2,m_3^2) \right) = \nonumber \\
 &&  - \frac{\pi}{p^2} \bigg\{
      c_1 {\rm K}\left(\kappa\right)
      + c_2 {\rm E}\left(\kappa\right) 
+ c_3 \Pi\left({q_{-+} \over q_{--}},
          \kappa\right) \nonumber \\
 && \;\;\;\;\;\;\;\; + c_4 \Pi\left(
             {{{(m_1-m_2)^2}\over {(m_1+m_2)^2}} {q_{-+} \over q_{--}}},
            \kappa\right)
           \;\bigg\} \nonumber \\
 && \;\;\;\;\;\; \times \Theta \left( p^2-(m_1+m_2+m_3)^2 \right) ,
     \label{ImT123} \\
 && \!\!\!\! \!\!\!\! c_1 =  
      4m_1m_2 \sqrt{q_{--} \over q_{++}} \nonumber \\
 &&  \;\;\; \times
       \Big\{ (p+m_3)^2-m_3 p +m_1 m_2 \Big\},
     \nonumber \\
 && \!\!\!\! \!\!\!\! 
    c_2 = {{m_1^2+m_2^2+m_3^2+p^2} \over 2} \sqrt{q_{++} q_{--}},
     \nonumber \\
 && \!\!\!\! \!\!\!\! c_3 = 
    {{8 m_1 m_2}
         \over \sqrt{q_{++} q_{--}}} 
     \Big\{(m_1^2+m_2^2)(p^2+m_3^2) \nonumber\\ 
 && \;\;\;\;\;\;\;\;\;\;\;\;\;\;\;\;\;\;\;\;\;
     -2m_1^2m_2^2-2m_3^2p^2 \Big\},
      \nonumber \\
 && \!\!\!\! \!\!\!\! 
   c_4 = - {{8m_1m_2(p^2-m_3^2)^2} \over \sqrt{q_{++} q_{--}}},
       \nonumber \\
 && \!\!\!\! \!\!\!\! \kappa^2 = \frac{q_{+-} q_{-+}}{q_{++} q_{--}}, \nonumber
\end{eqnarray}
with variables $q_{\pm\pm}$ corresponding to the physical and
unphysical thresholds
\begin{equation}
q_{\pm \pm}:=(p \pm m_3)^2-(m_1 \pm m_2)^2.
\end{equation}

This result is valid for all values of $m_1^2$,$m_2^2$ and $m_3^2$.
In special cases it leads to simpler formulae.
For equal masses one gets
\begin{eqnarray}
 &&\!\!\!\!\!\!\!\! Im(T_{123}(p^2;m^2,m^2,m^2)) \nonumber \\
 && \!\!\!\!  = - \frac{\pi}{p^2} \sqrt{(p-m)(p+3 m)} 
       \nonumber \\
 && \!\! \times
         \Big\{ 
            \frac{(p-m)(p^2+3 m^2)}{2} \, {\rm E}(\kappa) 
            -4 m^2 p \, {\rm K}(\kappa)
         \Big\}  
       \nonumber \\
  && \!\! \times \Theta(p^2 - 9 m^2) \, , \\
  && \!\!\!\!\!\!\!\! 
     \mbox{with} \;\;\; \kappa^2=\frac{(p+m)^3(p-3 m)}{(p-m)^3(p+3 m)},
\end{eqnarray}
involving only complete elliptic integrals of the first and second kind, i.e.
${_2 F_1}$ Gauss' hypergeometric functions.
If at least one mass is zero, $Im(T_{123})$ reduces to logarithms.

\boldmath
\subsection{The imaginary part of $T_{1234}$}
\unboldmath

The two-particle cut contribution to the discontinuity
of $T_{1234}$ was given in (\ref{DeltaT12342}),
\begin{eqnarray}
 && \!\!\!\!\!\! \Delta T^{(2)}_{1234}(p^2;m_1^2,m_2^2,m_3^2,m_4^2) 
     \nonumber \\
 && = \Delta B_0(p^2;m_1^2,m_2^2) \,
      B_0(m_1^2;m_3^2,m_4^2) \, .
 \label{DeltaT2}
\end{eqnarray}
As a product of a one-loop self-energy integral and a 
one-loop self-energy discontinuity it is
composed of elementary functions and gets a real part for
\begin{eqnarray}
 &&\!\!\!\!\!\!\!\! (m_3+m_4)^2 < m_1^2 \;\; \mbox{and} \;\;
                    (m_1+m_2)^2  < p^2 \,.
  \label{m1diff}
\end{eqnarray}

The three particle cut contribution looks very similar to that
of the London transport diagram.
Only one more propagator
$1/(t-m^2)$ has to be added in (\ref{ImT123short}).
The calculation yields
\begin{eqnarray}
 && \!\!\!\!\!\!\!\! \Delta T_{1234}^{(3)} (p^2;m_1^2,m_2^2,m_3^2,m_4^2) 
       \nonumber \\
 && \!\! = \frac{2\pi i}{p^2}
 \bigg\{ c_1 {\rm K}\left(\kappa\right) 
         + c_2 {\rm E}\left(\kappa\right)
         + c_3 \Pi\left({q_{-+} \over q_{--}},
          \kappa\right) 
      \nonumber \\
 && \;\;
       + c_4 \Pi\left({{{(m_3-m_4)^2}\over {(m_3+m_4)^2}} 
                      {q_{-+} \over q_{--}}},
                    \kappa\right)
       \nonumber \\
 && \;\; + c_5 \Pi\left({{{m_1^2-(m_3-m_4)^2}
                 \over {m_1^2-(m_3+m_4)^2}} {q_{-+} \over q_{--}}}
                   -i\epsilon,
            \kappa\right) 
        \bigg\}
    \nonumber \\
 &&  \times \Theta(p^2-(m_2+m_3+m_4)^2) \, ,
  \label{ImT1234} \\
 && \!\! c_1 = 4m_3m_4 \sqrt{q_{--} \over q_{++}}, \, 
         c_2 = \sqrt{q_{++} q_{--}}, \nonumber \\
 && \!\! c_3 = 
     {{8 m_3 m_4 (p^2-m_1^2+m_2^2+m_3^2+m_4^2)}
            \over \sqrt{q_{++} q_{--}}},  \nonumber \\
 && \!\! c_4 =
      - {{8m_3m_4(p^2-m_2^2)^2} \over {m_1^2\sqrt{q_{++} q_{--}}}},
       \nonumber \\ 
 && \!\! c_5= {{8m_3m_4\lambda(p^2,m_1^2,m_2^2)} 
                 \over {m_1^2\sqrt{q_{++} q_{--}}}}, \nonumber \\
 && \!\! \kappa^2 = \frac{q_{+-} q_{-+}}{q_{++} q_{--}}, \nonumber
\end{eqnarray}
with
\begin{equation}
q_{\pm \pm}:=(p \pm m_2)^2-(m_3 \pm m_4)^2.
\end{equation}

In the case (\ref{m1diff}) the characteristic c of the last $\Pi$-function
in (\ref{ImT1234}) is greater than $1$,
\begin{eqnarray}
  && c= {{{m_1^2-(m_3-m_4)^2}
          \over {m_1^2-(m_3+m_4)^2}} {q_{-+} \over q_{--}}} > 1 \, ,
\end{eqnarray}
which requires an analytic continuation of that function.
A comprehensive discussion of the analytic properties of
the elliptic integrals can be found in \cite{Tricomi}.
The $i \epsilon$-prescription in (\ref{ImT1234})
ensures that $\Delta T^{(3)}_{1234}$ gets the correct
real part, given through
\begin{eqnarray}
  &&  Im \left( \Pi (c-i\epsilon,\kappa) \right) \nonumber \\
  && \;\;\;\;\;
   = {1 \over {2 i}} \left( \Pi(c-i\epsilon,\kappa)-\Pi(c+i\epsilon,\kappa)
             \right)
     \nonumber \\
  && \;\;\;\;\;
     = - \frac{ \pi}{2} \sqrt\frac{c}{(c-1)(c-\kappa^2)} \,. 
\end{eqnarray}
This contribution cancels the real part of $\Delta T^{(2)}_{1234}$.
Consequently  $\Delta T_{1234}$ is always purely imaginary.

Numerical checks show the agreement of the results of (\ref{ImT123}) for
$Im(T_{123})$ and of
\begin{eqnarray}
  Im(T_{1234})=\frac{1}{2 i} \left( \Delta T^{(2)}_{1234}
  + \Delta T^{(3)}_{1234} \right)
\end{eqnarray}
with previously published tables \cite{Buza,Berends}.

\section{ONE-DIMENSIONAL INTEGRAL REPRESENTATIONS} 

\subsection{A general approach to two-loop integrals containing
            a self-energy subloop}

An alternative method to the series expansion of the two-loop scalar
diagrams consists in the derivation of one-dimensional integral
representations. These are built up from one-loop self-energy functions
$B_0$ coming from the self-energy subloop and the remaining one-loop
integral. They can be derived by using a dispersion representation of
the $B_0$ function.

A two-loop diagram with only three-vertices
\vskip3mm
\beginpicture

\setcoordinatesystem units <.013mm,.013mm>
\setplotsymbol (.)

\circulararc 300 degrees from 866 500 center at 0 0
\circulararc 60 degrees from -866 -500 center at -1732 0

\plot 0 1000   0 1500 /
\plot 0 -1000   0 -1500 /
\plot 866 500  1299 750 /
\plot 866 -500  1299 -750 /

\put{$p_1$} [lb] at 50 -1500 
\put{$p_2$} [lt] at 1350 -750
\put{$p_{N-1}$} [lt] at 1350 750
\put{$p_{N}$} [lb] at 50 1400
\put{$m_{N+3}$} [rt] at -500 -866
\put{$m_1$} [rb] at 500 -866
\put{$m_{N-1}$} [lb] at 300 900
\put{$m_{N+2}$} [rb] at -500 866
\put{$m_{N}$} [lB] at -700  0
\put{$m_{N+1}$} [rB] at -1100 0
\put{,} at 1500 0
\put{ } [lB] at -2500 0

\setdots
\circulararc 60 degrees from 866 -500 center at 0 0

\setplotsymbol ({\circle*{5}} [Bl])

\plot 866 500  866 501 /
\plot 866 -500  866 -501 /
\plot 0 1000  0 1001 /
\plot 0 -1000  0 -1001 /
\plot -866 500  -866 501 /
\plot -866 -500  -866 -501 /

\setplotsymbol (.)
\setsolid
\endpicture
\vskip3mm
\noindent
where $k$ is the momentum flowing through the self-energy insertion,
can in a first step be reduced to simpler diagrams by a
decomposition into partial fractions
\begin{eqnarray}
 && \!\!\!\!\!\!\!\! \frac{1}{k^2-m_{N+2}^2} \frac{1}{k^2-m_{N+3}^2} =
   \frac{1}{m_{N+2}^2-m_{N+3}^2} \nonumber \\
 && \;\;\;\;\;\;\;\; \times  
     \left( \frac{1}{k^2-m_{N+2}^2} - \frac{1}{k^2-m_{N+3}^2} \right).
   \nonumber
\end{eqnarray}
This results for the diagram in
\begin{eqnarray}
  && \!\!\!\! \!\!\!\! \! T_{1 \ldots N+3}(p_i;m_i^2) = 
    \frac{1}{m_{N+2}^2-m_{N+3}^2} \nonumber \\
  && \!\! \times \bigg(
       T_{1 \ldots N+2}(p_i;m_1^2,\ldots,m_{N+1}^2,m_{N+2}^2)
   \nonumber \\
  && \;\;  - T_{1 \ldots N+2}(p_i;m_1^2,\ldots,m_{N+1}^2,m_{N+3}^2)
    \bigg).
\end{eqnarray}
The difference has to be replaced by a derivative if $m_{N+2}^2=m_{N+3}^2$.

Insertion of the dispersion representation for the self-energy
subloop leads to
\begin{eqnarray}
  && \!\!\!\! \!\!\!\! 
       T_{1 \ldots N+2}(p_i;m_1^2,\ldots,m_{N+1}^2,m_{N+2}^2) \nonumber \\
  && \!\!\!\! = \int d^D k \, B_0(k^2;m_N^2,m_{N+1}^2) 
                \frac{1}{(k+p_1)^2-m_1^2}  \nonumber \\
  &&        \ldots\times \frac{1}{(k+p_1+\dots+p_{N-1})^2-m_{N-1}^2} 
                \nonumber \\
  && \times     \frac{1}{k^2-m_{N+2}^2} \nonumber \\
  && \!\!\!\! = \frac{1}{2 \pi i} \int\limits_{s_0}^\infty ds \,
                 \Delta B_0(s;m_N^2,m_{N+1}^2) \nonumber \\
  && \times      \int d^D k \,
                 \frac{-1}{k^2-s+i\epsilon}
                 \frac{1}{k^2-m_{N+2}^2} \nonumber \\
  && \times     \frac{1}{(k+p_1)^2-m_1^2} \ldots \nonumber \\
  && \times     \frac{1}{(k+p_1+\dots+p_{N-1})^2-m_{N-1}^2}.
\end{eqnarray}
After a further decomposition into partial fractions,
\begin{eqnarray}
  && \!\!\!\!\!\!\!\! \frac{-1}{k^2-s} \;
      \frac{1}{k^2-m_{N+2}^2} \nonumber \\ 
  && \!\!\!\! = \frac{1}{s-m_{N+2}^2} 
                \left(
                \frac{1}{k^2-m_{N+2}^2} - \frac{1}{k^2-s} \right),
\end{eqnarray}
the $k$-integrations and one of the $s$-integrations
can be performed yielding
\begin{eqnarray}
  && \!\!\!\! \!\!\!\! 
      T_{1 \ldots N+2}(p_i;m_1^2,\ldots,m_{N+1}^2,m_{N+2}^2) \nonumber \\
  && \!\!\!\! = B_0(m_{N+2}^2;m_N^2,m_{N+1}^2) \nonumber \\
  && \;\;\; \times T^{(1)}(p_i;m_1^2,\ldots,m_{N-1}^2,m_{N+2}^2) \nonumber \\
  && - \frac{1}{2 \pi i} \int\limits_{s_0}^\infty ds \,
       \frac{\Delta B_0(s;m_N^2,m_{N+1}^2)}{s-m_{N+2}^2} \nonumber \\
  && \;\;\; \times T^{(1)}(p_i;m_1^2,\ldots,m_{N-1}^2,s).
    \label{1disp}
\end{eqnarray}
$T^{(1)}$ denotes a one-loop N-point function in which $s$ enters
in the remaining one-dimensional integration as a mass variable.
%

A diagram with two four-vertices leads to a result which is
similar to the remaining integration in (\ref{1disp}),
\vskip3mm
\beginpicture

\setcoordinatesystem units <.01mm,.01mm>
\setplotsymbol (.)

\circulararc 300 degrees from 866 500 center at 0 0
\circulararc 60 degrees from -866 -500 center at -1732 0

\plot 866 500  1299 750 /
\plot 866 -500  1299 -750 /
\plot -866 500 -1299 750 /
\plot -866 -500 -1299 -750 /

\put{$p_1$} [lt] at -1250 -750 
\put{$p_2$} [lt] at 1350 -750
\put{$p_{N-1}$} [lb] at 1350 750
\put{$p_{N}$} [lb] at -1250 750
\put{$m_1$} [lb] at -50 -900
\put{$m_{N-1}$} [lb] at -50 1000
\put{$m_{N}$} [lB] at -700  0
\put{$m_{N+1}$} [rB] at -1100 0

\put{ } [Bl] at -3500 0

\setdots
\circulararc 60 degrees from 866 -500 center at 0 0

\setplotsymbol ({\circle*{5}} [Bl])

\plot 866 500  866 501 /
\plot 866 -500  866 -501 /
\plot -866 500  -866 501 /
\plot -866 -500  -866 -501 /

\endpicture
\vskip3mm
\begin{eqnarray}
 \!\!\!\! \!\!\!\! &&
   T_{1 \ldots N+1}(p_i;m_i^2) \nonumber \\
 && \!\!\!\! = \frac{1}{2\pi i} 
        \int \limits_{s_0}^\infty ds\, \Delta B_0(s;m_{N}^2,m_{N+1}^2)
    \nonumber \\
 && \times
   \int d^D k \frac{-1}{k^2-s}
              \frac{1}{(k+p_1)^2-m_1^2} \ldots \nonumber \\
 && \times \frac{1}{(k+p_1+\ldots+p_{N-1})^2-m_{N-1}^2}
         \nonumber \\
 && \!\!\!\! 
    =  - \frac{1}{2\pi i} \int \limits_{s_0}^\infty ds\,
        \Delta B_0(s;m_{N}^2,m_{N+1}^2) \nonumber \\
 && \times T^{(1)}(p_i;m_1^2,\ldots,m_{N-1}^2,s) \, .
 \label{2disp}
\end{eqnarray}
As an example, we apply (\ref{1disp}) to $T_{1234}$:
\begin{eqnarray}
  && \!\!\!\!\!
     T_{1234}(p^2;m_1^2,m_2^2,m_3^2,m_4^2) \nonumber \\
  && =  B_0(m_1^2;m_3^2,m_4^2) B_0(p^2;m_1^2,m_2^2)  \nonumber \\
  &&\;\;\;\; - \frac{1}{2\pi i} \int \limits_{(m_3+m_4)^2}^\infty ds\,
       \frac{\Delta B_0(s;m_3^2,m_4^2)}{s-m_1^2+i\epsilon} \nonumber \\
  && \;\;\;\;\;\;\;\;\;\;\;\;\; \times
        B_0(p^2;s,m_2^2).
  \label{T1234disp}
\end{eqnarray}
An application of (\ref{2disp}) to the London transport diagram
leads to
\begin{eqnarray}
  && T_{123} = - \frac{1}{2\pi i} \int \limits_{(m_2+m_3)^2}^\infty
   \!\!\!\! ds\,
        \Delta B_0(s;m_2^2,m_3^2) \nonumber \\
  && \;\;\;\;\;\;\;\;\;\;\;\;\; \times B_0(p^2;s,m_1^2),
  \label{T123disp}
\end{eqnarray}
a result which would also follow from (\ref{disp}).

The representations (\ref{T123disp}) and (\ref{T1234disp}) with the
subtractions (\ref{DefT123N}) and (\ref{finite}) provide efficient ways to
calculate
$T_{123N}$ and $T_{1234N}$ in all parameter regions. The results
agree numerically with those published in \cite{Buza,Berends}.

\subsection{Integral representation of the master diagram}

In this section a one-dimensional integral representation
for the master diagram $T_{12345}$ (fig.~\ref{fig:top})
in the general mass case is
derived. This representation uses only elementary functions
and is thus well suited for numerical evaluation.

We use the dispersion relation. The two-particle cut contributions are
denoted $T^{(2a)}$ for
the cut through the propagators 1 and 2, and $T^{(2b)}$ for
the cut 4-5.
The Cutkosky rules yield
\begin{eqnarray}
  &&\!\! \Delta T^{(2a)}(p^2;m_1^2,m_2^2,m_3^2,m_4^2,m_5^2) \nonumber \\
  && \;\; =  \Delta B_0(p^2;m_1^2,m_2^2) \nonumber \\
  && \;\;\;\; \times
          \left( C_0 (p^2,m_1^2,m_2^2; m_4^2,m_3^2, m_5^2) \right)^*.
   \label{DT2}
\end{eqnarray}
This shows that the two-particle cut contributions are given by
the self-energy discontinuity and the complex conjugate of the
triangle diagram $C_0$.
For a further evaluation of (\ref{DT2}) one can introduce the dispersion
representation of $C_0$,
\begin{eqnarray}
  && \!\!\!\!\!\!\!\! C_0(p_1^2,p_2^2,p_3^2;m_1^2,m_2^2,m_3^2) \nonumber \\
  && \!\!\!\! =  \frac{1}{2 \pi i}
      \int\limits_{(m_1+m_2)^2}^\infty \!\!\! dt \, 
       \frac{\Delta C_0 (t)}{t-p_1^2-i\epsilon}
       + C_{0an} \, , \label{C0disp}
\end{eqnarray}
with a contribution of the discontinuity belonging to the normal threshold
\begin{eqnarray}
  && \Delta C_0(t,p_2^2,p_3^2;m_1^2,m_2^2,m_3^2) \nonumber \\
  && \;\; = - \frac{2 \pi i}{\sqrt{\lambda(t,p_2^2,p_3^2)}}
                 \log\frac{a+b}{a-b} \label{DeltaC0} \\
  && a=t(t + 2 m_3^2 - p_2^2 -p_3^2 - m_1^2 - m_2^2) \nonumber \\
  && \;\;\;\;\;\; + (p_3^2-p_2^2) (m_1^2-m_2^2),
     \nonumber \\
  && b={\sqrt{\lambda(t,p_2^2,p_3^2)}}{\sqrt{\lambda(t,m_1^2,m_2^2)}},
     \nonumber
\end{eqnarray}
and a contribution $C_{0an}$ belonging to the anomalous threshold,
resulting from the leading Landau singularity. This anomalous
threshold at
\begin{eqnarray}
   t_1 &=& m_3^2(-m_3^2+p_2^2+p_3^2+m_1^2+m_2^2) \nonumber \\
       && - (p_2^2-m_2^2)(p_3^2-m_1^2)
               \nonumber \\
       && - \sqrt{\lambda(p_2^2,m_2^2,m_3^2)} \sqrt{\lambda(p_3^2,m_1^2,m_3^2)}
\end{eqnarray}
occurs if
\begin{eqnarray}
  && m_1 p_2^2 + m_2 p_3^2 - m_3^2 (m_1+m_2) \nonumber \\
  && - m_1 m_2 (m_1+m_2) > 0.
     \label{AnormThrCond}
\end{eqnarray}
Following the argument outlined in \cite{Barton} one can show that
its contribution is
\begin{eqnarray}
  && \!\!\!\! \!\!\!\! 
     C_{0an} = \!\!\!\! \int\limits_{(m_1+m_2)^2} ^{t_1}
       \frac{dt}{t-p_1^2} \;
       \frac{ 2 \pi i}{\sqrt{\lambda(t,p_2^2,p_3^2)}}
\end{eqnarray}
and yields logarithms and square roots. In those cases, where $t_1$
is situated
on the real axis and $t_1 > t_0=(m_1+m_2)^2 $, this contribution can also be
taken into account by choosing the appropriate sheets for the
logarithm in (\ref{DeltaC0}).

After insertion of (\ref{C0disp}) into (\ref{DT2})
the integrations in the
dispersion integral can be interchanged and lead to
\begin{eqnarray}
  && \!\!\!\! \!\!\!\!
      T^{(2a)}(p^2;m_1^2,m_2^2,m_3^2,m_4^2,m_5^2) \nonumber \\
  && \!\!\!\! = \frac{1}{2 \pi i} \int\limits_{s_0}^\infty 
           \frac{ds}{s-p^2-i\epsilon}
           \Delta T^{(2a)}(s;m_i^2) \nonumber \\
  && \!\!\!\! = \left( \frac{1}{2 \pi i} \right)^2 \int\limits_{t_0}^\infty dt
        \int\limits_{s_0}^\infty ds \nonumber \\
  && \;\;\;\;\;\; \times
        \frac{\Delta C_0(t)\Delta B_0 (s)}
             {(s-p^2-i\epsilon)(t-s + i\epsilon)}
        \nonumber \\
  && \;\; + \frac{1}{2\pi i} \int\limits_{s_0}^\infty ds \,
               \frac{\Delta B_0(s;m_1^2,m_2^2)}
                    {s-p^2-i\epsilon} \nonumber \\
  && \;\;\;\;\;\; \times
          \left( C_{0an} (s,m_1^2,m_2^2;m_5^2,m_4^2,m_3^2) \right)^* 
      \nonumber \\
  && \!\!\!\!
     =  \frac{1}{2 \pi i} \int\limits_{(m_4+m_5)^2}^\infty \!\!\!\! dt 
        \frac{\Delta C_0 (t,m_1^2,m_2^2; m_4^2,m_3^2, m_5^2)}{t-p^2-i\epsilon}
      \nonumber \\
  && \;\;\;\;\;\; \times
        \left(B_0(p^2;m_1^2,m_2^2) - B_0 (t;m_1^2,m_2^2) \right) 
        \nonumber \\
  && \;\;\; + \frac{1}{2\pi i} \int\limits_{(m_1+m_2)^2}^\infty ds \,
               \frac{\Delta B_0(s;m_1^2,m_2^2)}
                    {s-p^2-i\epsilon} \nonumber \\
  && \;\;\;\;\;\; \times
        \left(C_{0an} (s,m_1^2,m_2^2;m_5^2,m_4^2,m_3^2) \right) ^* . 
    \label{T2}
\end{eqnarray}
To calculate the contribution of the other two-particle cut
$\Delta T^{(2b)}$, the masses have to be interchanged
according to
$m_1 \leftrightarrow m_4$, $m_2 \leftrightarrow m_5$. Furthermore
no complex conjugated form of $C_0$ has to be inserted in (\ref{DT2}),
resulting in a complex conjugated form of $B_0(t;m_4^2,m_5^2)$ in
the first term of
(\ref{T2}) and no complex conjugated form of $C_{0an}$
in the second term of (\ref{T2}).

To evaluate the three-particle cut contributions,
we keep close to Broadhurst's approach \cite{Broadhurst}.
The contribution $T^{(3a)}$ refers to the cut
$2$-$3$-$4$, $T^{(3b)}$ refers to the cut $1$-$3$-$5$.
The Cutkosky rules yield
\begin{eqnarray}
  && \!\! \Delta T^{(3a)}(p^2;m_1^2,m_2^2,m_3^2,m_4^2,m_5^2) \nonumber \\
  && \;\; =  - \frac{1}{2 \pi i} 
      \int\limits_{(m_3+m_4)^2}^{(\sqrt{s}-m_2)^2} \!\! dt \,
     \frac{\Delta B_0(p^2;t,m_2^2)}
          {t-m_1^2+i\epsilon} \nonumber \\
  && \;\;\;\; \times
      \Delta C_0(t,p^2,m_2^2;m_3^2,m_4^2,m_5^2) . \label{DT3}
\end{eqnarray}
A partial integration in
(\ref{DT3}) leads to a form which simplifies the
dispersion integral considerably
\begin{eqnarray}
 && \!\!\!\!\! T^{(3a)}(p^2;m_i^2) = \nonumber \\
 &&  = \int\limits_{s_0}^\infty ds \,
    \frac{\Delta T^{(3a)} (s;m_i^2)}
         {s-p^2-i\epsilon} \nonumber \\
 &&  =
     \int\limits_{(m_2+m_3+m_4)^2}^\infty \frac{ds}{s(s-p^2)}
        \nonumber \\
  && \;\;\;\;\; \times 
     \int\limits_{(m_3+m_4)^2}^{(\sqrt{s}-m_2)^2} \!\! dt \, 
      \log \left( \frac{t}{m_1^2}-1+i\epsilon \right) \nonumber \\
  && \;\;\;\;\; \times
        \sqrt{\lambda(s,t,m_2^2)} \nonumber \\
  && \;\;\;\;\; \times R(s,t,m_2^2,m_3^2,m_4^2,m_5^2)
           \label{T3partint} \, ,
\end{eqnarray}
where $R$ is for the case of different masses a rational function
of $s$ containing only first order poles.
An interchange of integrations,
\begin{eqnarray}
  && \int\limits_{(m_2+m_3+m_4)^2}^\infty ds
        \int\limits_{(m_3+m_4)^2}^{(\sqrt{s}-m_2)^2} dt \nonumber \\
  && \;\; =
     \int\limits_{(m_3+m_4)^2}^\infty dt 
     \int\limits_{(m_2+\sqrt{t})^2}^\infty ds,
\end{eqnarray}
and decomposition of $R(s)$ into partial fractions leads
to integrations of the type
\begin{eqnarray}
  && \int\limits _{(\sqrt{t}+m_2)^2}^\infty \frac{ds}{s-s_i-i\epsilon}
             \frac{\sqrt{\lambda(s,t,m_2^2)}}{s} \nonumber \\
  && \;\;\;\; =  B_0(s_i;t,m_2^2),
\end{eqnarray}
i.e. one-loop self-energy integrals. In the final expression all
ultraviolet divergences of these $B_0$-functions cancel.
The result for the three-particle cut contribution of the master
diagram is then
\begin{eqnarray}
  &&\!\!\!\!\! T^{(3a)}(p^2;m_i^2) \nonumber \\
  &&\!\!  = \int\limits_{(m_3+m_4)^2}^\infty dt \;
        \log \left( \frac{t}{m_1^2}-1+i\epsilon \right)
     \nonumber \\
  && \; \times
        \frac{\sqrt{\lambda(t,m_4^2,m_3^2)}}{t\, m_3^2} \nonumber \\
  && \;
      \times \sum_{i=1}^4 c_i 
        \frac{B_0(p^2;t,m_2^2)-B_0(s_i;t,m_2^2)} {s_i^2-p^2},
       \label{T12345-3} \\
  && \!\!\!\!\! \mbox{with} \nonumber \\
  &&\!\!\!\!\! s_{1/2}=
           \frac{t+m_2^2+m_4^2+m_5^2-m_3^2}{2} \nonumber \\
  && \!\! +   \frac{(t-m_4^2)(m_5^2-m_2^2)}
                   {2m_3^2} \nonumber \\
  && \!\!
          \pm \frac{\sqrt{\lambda(t,m_3^2,m_4^2)\lambda(m_2^2,m_3^2,m_5^2)}}
                   {2m_3^2}, \nonumber \\
  && \!\!\!\!\! s_{3/4}=(m_2 \pm \sqrt{t})^2, \nonumber \\
  && \!\!\!\!\! r_1=t(2m_2^2-m_5^2+m_3^2)-m_2^2(m_4^2-m_3^2), \nonumber \\
  && \!\!\!\!\! r_2=(t-m_2^2)(t(m_5^2-m_3^2)-m_2^2(m_4^2-m_3^2)), \nonumber \\
  && \!\!\!\!\! r_3=\frac{m_3^2(m_3^2-t-m_4^2)}
                         {\lambda(t,m_4^2,m_3^2)}, \nonumber \\
  && \!\!\!\!\! r_4=m_3^2 \nonumber \\
  && \!\! + \bigg\{t(2m_4^2m_3^2-m_4^2m_2^2+m_5^2m_4^2+m_5^2m_3^2)
            \nonumber \\
  &&        + m_4^2(m_2^2-m_5^2)(m_4^2-m_3^2) \nonumber \\
  &&        + m_5^2 m_3^2 (m_4^2-m_3^2) 
            \bigg\} \frac{1}
                        {\lambda(t,m_4^2,m_3^2)}, \nonumber \\
  && \!\!\!\!\! c_1= \bigg\{s_1 r_1 +r_2 \nonumber \\
  && \;\;\;  + s_1(s_1^2-(s_3+s_4)-s_3 s_4)(s_1 r_3- r_4) \bigg\} \nonumber \\
  && \;\times \frac{1}{\prod\limits_{i=2}^4 (s_1-s_i)}, 
     \;\; c_2=c_1(s_1 \leftrightarrow s_2), \nonumber\\
  && \!\!\!\!\! c_4=\frac{s_4 r_1 + r_2}{\prod\limits _{i=1}^3 (s_4-s_i)}, \;\;
     c_3=c_4(s_3 \leftrightarrow s_4). \nonumber
\end{eqnarray}

Some special mass cases have to be considered:
double poles occur in $R(s)$ if $m_2=0$ or
if $\lambda(m_2^2,m_5^2,m_3^2)=0$. The decomposition into partial fractions
leads then to modified results involving functions
\begin{eqnarray}
  && \int\limits _{(\sqrt{t}+m_2)^2}^\infty \frac{ds}{(s-s_i^2-i\epsilon)^2}
             \frac{\sqrt{\lambda(s,t,m_2^2)}}{s} \nonumber \\
  && \;\;\;\;\;\; =
      \frac{\partial B_0(s_i,t,m_2^2)}{\partial s_i}.
\end{eqnarray}
Another modification of the decomposition into partial fractions
occurs if $m_3=0$, in which case the result is considerably simpler.

Broadhurst has evaluated the master diagram for some cases
of physical interest \cite{Broadhurst}. They
all belong to the special cases mentioned above, if the symmetries
of the master diagram are exploited.
The formulae (\ref{T2}) and (\ref{T12345-3})
provide a method to calculate the master diagram in the general
mass case. The occurring elementary one-dimensional integrations
are numerically
stable and the results agree with those of Kreimer's method
\cite{Berends}. Compared with the latter
the one-dimensional integral representation is much faster,
especially if high accuracy is required.

\section{CONCLUSIONS}
On the long way to complete two-loop calculations in the 
Standard Model this paper focuses on the self-energy diagrams. 

When one restricts calculations to self-energies, these should
be useful as building blocks with suitable theoretical properties. In
particular, their gauge dependence has to be considered.
We have discussed this problem for light fermionic
contributions to the two-loop $Z$ \se\ by applying two different
approaches, the pinch technique (PT) and the background-field
method (BFM). The PT aims on eliminating the gauge parameter
dependence of Green functions within $R_{\xi}$-gauges, while
in the BFM Green functions are derived from a gauge invariant
effective action. Whereas the BFM is valid to all orders in
perturbation theory, the PT had so far been restricted to the
one-loop level. 

In this paper an extension of the PT to the
two-loop case has been worked out yielding (within $R_{\xi}$-gauges)
a gauge parameter independent result. It involves a rearrangement
of contributions between different Green functions and between 
one-particle irreducible and reducible diagrams.
The BFM has been shown to provide a more general framework.
Like at one-loop order, it
incorporates the PT results as a special case. The BFM vertex
functions 
fulfill simple Ward identities which
are a direct consequence of gauge invariance. This holds for all
values of the quantum gauge parameter $\xi_Q$.

For large parts of the actual calculation of the two-loop 
\se\ diagrams we use an algebraic approach allowing for a high
degree of automatization. By applying a method for tensor integral
decomposition, all two-loop \ses\ are reduced to a minimal
basis of standard scalar integrals. 

The evaluation of the scalar diagrams is the remaining problem 
and is in 
essence more involved than in the case of one-loop diagrams.
The reason is that for the general mass case needed in the SM
functions beyond the usual (poly) logarithms 
are required.
Analytic and numerical approaches have been discussed.

We have derived analytic results 
in terms of generalized hypergeometric functions for the scalar
\se\ integrals
with three and four propagators and arbitrary masses.
This offers the possibility to use well-known mathematical 
techniques like
analytic continuation, partial differential equations, or contour 
integral representations.
The known formulas for integrals with vanishing masses are 
obtained as special cases and consequently are unified 
in one result in $D$ dimensions.
For the general mass case of the master diagram a representation 
in the form of a generalized hypergeometric function is not yet
known. 

When one is interested only in the imaginary parts, 
an alternative analytic
result in four dimensions is obtained in terms of 
complete elliptic integrals.
Since their properties are well-known they are easily 
accessible for numerical evaluations.

Finally, we have derived a one-dimensional integral representation 
for all two-loop diagrams containing a self-energy insertion.
For the two-loop self-energy diagrams treated in this paper,
the integrand is
composed of elementary functions only and the representation 
is valid for all values of $p^2$.
Also for the master diagram a one-dimensional elementary integral 
representation is derived. The main application of these integrals 
is for numerical evaluations, giving a good alternative to the 
existing two-dimensional integrals. 

Once adequate techniques for the self-energy diagrams are available, 
one could envisage practical applications for physics predictions.
For the electroweak theory the obvious application is to the 
gauge boson self-energies which play a role in the
$M_W\,-\,M_Z$ mass relation and details of the $Z$ line shape.
For QED the two-loop vacuum polarization is known since a long time 
\cite{Sabry}, but the electron two-loop 
self-energy was never fully calculated.



\section*{ACKNOWLEDGEMENTS}

The authors would like to thank the organizers of the Teupitz
meeting for the kind hospitality during the workshop. We also
acknowledge fruitful discussions with our colleagues, especially
with A.~Denner, S.~Dittmaier, R.~Scharf and J.B.~Tausk.

\begin{sloppypar}

\end{sloppypar}
\end{document}